\def\frac#1#2{{{{#1}}\over{{#2}}}}

\documentclass[12pt]{article}
\usepackage{colordvi}
\usepackage{amsmath, amssymb, graphics}

\newsavebox{\ns}
\newsavebox{\dbrane}
\newsavebox{\dbshort}

\usepackage{epsfig}
\usepackage{amssymb}

\def\appendix{{\newpage\section*{Appendix}}\let\appendix\section%
        {\setcounter{section}{0}
        \gdef\thesection{\Alph{section}}}\section}

\newcommand\ba{\begin{eqnarray}}
\newcommand\ea{\end{eqnarray}}
\newcommand\be{\begin{equation}}
\newcommand\ee{\end{equation}}

\newcommand{\nn}{\nonumber}

\def\Dslash{\,\,{\raise.15ex\hbox{/}\mkern-12mu D}}
\def\Dbarslash{\,\,{\raise.15ex\hbox{/}\mkern-12mu {\bar D}}}
\def\delslash{\,\,{\raise.15ex\hbox{/}\mkern-9mu \partial}}
\def\delbarslash{\,\,{\raise.15ex\hbox{/}\mkern-9mu {\bar\partial}}}
\def\pslash{\,\,{\raise.15ex\hbox{/}\mkern-9mu p}}
\def\calDslash{\,\,{\raise.15ex\hbox{/}\mkern-12mu {\cal D}}}

\newcommand{\hh}{{1\over 2}}

\newcommand{\<}{ <  \hskip -.05in < } % heer lessthan problem
\renewcommand{\ggg}{ >  \hskip -.05in > } % heer lessthan problem
\renewcommand{\ll}{_}
\newcommand{\uu}{^}
\newcommand{\pp}{\partial}

\renewcommand{\exp}[1]{{\rm exp}\left \{#1 \right \}}

\newcommand{\m}{\mu}

\renewcommand{\m}{\mu}

\newcommand{\s}{\sigma}
\renewcommand{\t}{\tau}
\newcommand{\G}{\Gamma}
\newcommand{\g}{\gamma}

\newcommand{\sqd}{^2}
\newcommand{\zb}{{\bar{z}}}

\renewcommand{\hh}{{1\over 2}}

\newcommand{\eee}[1]{\ba{#1}\ea}
\renewcommand{\th}{\theta}
\renewcommand{\t}{\tau}

\newcommand{\llsk}{\hskip .5in}

\newcommand{\apr}{{\alpha^\prime} {}}

\newcommand{\IZ}{\relax\ifmmode\mathchoice
{\hbox{\cmss Z\kern-.4em Z}}{\hbox{\cmss Z\kern-.4em Z}}
{\lower.9pt\hbox{\cmsss Z\kern-.4em Z}} {\lower1.2pt\hbox{\cmsss
Z\kern-.4em Z}}\else{\cmss Z\kern-.4em Z}\fi} \font\cmss=cmss10
\font\cmsss=cmss10 at 7pt
\newcommand{\inbar}{\,\vrule height1.5ex width.4pt depth0pt}
\newcommand{\IC}{{\relax\hbox{$\inbar\kern-.3em{\rm C}$}}}
\newcommand{\IQ}{{\relax\hbox{$\inbar\kern-.3em{\rm Q}$}}}
\newcommand{\IH}{{\relax\hbox{$\inbar\kern-.3em{\rm H}$}}}
\newcommand{\IP}{\relax{\rm I\kern-.18em P}}

\newcommand{\ed}{\dot{e}}

\newcommand{\cc}{{\cal C}}

\renewcommand{\cc}{{c_1}}

\renewcommand{\cc}{c}

\newcommand{\IR}{\relax{\rm I\kern-.18em R}}
\def\blfootnote{\xdef\@thefnmark{}\@footnotetext}

\renewcommand{\cc}[1]{\cite{#1}}

\newcommand{\bm}{\begin{matrix}}
\renewcommand{\em}{\end{matrix}}

\def\lrdd{\left (}
\def\rrdd{\right )}
\def\lsqq{\left [}
\def\rsqq{\right ]}
\def\exp#1{{\rm exp}\left ( {#1} \right )}

\newcommand{\upp}[1]{^{({#1})}{}}

\newcommand{\co}{{\cal O}}

\newcommand{\rr}[1]{(\ref{{#1}})}
\newcommand{\bbb}{\ba\begin{array}{c}}
\renewcommand{\eee}{\nonumber\end{array}\ea}
\newcommand{\een}[1]{\label{#1}\end{array}\ea}

\def\hilo{{}_{{}_{{}_{{}_{{}_{}}}}} {}^{{}^{{}^{}}}}

\newcommand{\heading}[1]{\begin{center}\it {#1} \rm \end{center}}

\def\lrdd{\left ( ~}
\def\rrdd{\hilo \right )}
\def\lsqq{\left [ ~}
\def\rsqq{\hilo \right ]}

\def\bi{\begin{itemize}}
\def\ei{\end{itemize}}

\def\ed{\end{document}}

\def\cc{{\cal C}}

\renewcommand{\rr}[1]{(\ref{#1})}

\def\cc{\,}

\def\xxn{\\ \\}
\def\xxx{\nn\\ \nn\\}

\newcommand{\aaa}[1]{}

\def\lleq{< \hskip -.08in <}

\newcommand{\lp}[1]{_{({#1})}}
\def\bfr{{\tt r}}
\def\bfrzero{{\tt r}_0}

\def\pmacro{\tilde{p}}
\def\qmacro{\tilde{q}}
\def\pstmacro{\tilde{p}^*}
\def\qstmacro{\tilde{q}^*}

\begin{document}

\begin{titlepage}
\begin{flushright}
%PREPRINT-NUMBER\\
IPMU-10-0156\\
\end{flushright}
\vspace{15 mm}
\begin{center}
  {\Large \bf  Dynamical Cobordisms   \\ in \\ \vspace{2 mm} General Relativity and String Theory
%\\ \vspace{.11in}
%to the duality web
}
\end{center}
\vspace{6 mm}
\begin{center}
{ Simeon Hellerman$^1$ and Matthew Kleban$^2$ }\\
\vspace{6mm}
{\it $^1$Institute for the Physics and Mathematics of the Universe\\
The University of Tokyo \\
 Kashiwa, Chiba  277-8582, Japan\\}
 \vspace{6mm}
{\it $^2$Center for Cosmology and Particle Physics\\
New York University \\
4 Washington Place\\
New York, NY 10003
 }

\end{center}
\vspace{6 mm}
\begin{center}
{\large Abstract}
\end{center}
\noindent
%%%%%%%%%%%%%%%%%%%%%%%%%%%%%%%%%%%%%%%%%%%%%%%%%%%%%%%%%%%%%%%%%%%%%%%%%%%%%%%%%%%%

We describe a class of time-dependent solutions in string- or M-theory that are exact with respect to $\apr$ and curvature corrections and interpolate in physical space between regions in which the low energy physics is well-approximated by different string theories and string compactifications.  The regions are connected by expanding ``domain walls" but are not separated by causal horizons, and physical excitations can propagate between them.  As specific examples we construct solutions that interpolate between oriented and unoriented string theories, and also between type II and heterotic theories.  Our solutions can be weakly curved and under perturbative control everywhere and can asymptote to supersymmetric at late times.

%%%%%%%%%%%%%%%%%%%%%%%%%%%%%%%%%%%%%%%%%%%%%%%%%%%%%%%%%%%%%%%%%%%%%%%%%%%%%%%%%%%%%
\vspace{1cm}
\begin{flushleft}
\today
%September 16, 2007
\end{flushleft}
\end{titlepage}
\tableofcontents
\newpage

% BEGIN CUT

\section{Introduction}

The great majority of research in string theory has focused on supersymmetric, time-independent configurations.  Such solutions generally do not admit domain walls that interpolate between different string theories.
However, when one relaxes these strictures even slightly, a much larger range of solutions becomes available.  For example \cite{hs1,hs2,hs3,hs4,hs5,hs6}
demonstrated that non-supersymmetric string theories are dynamically connected to supersymmetric vacua by tachyon condensation.  In \cite{klebanexpand}, one of us showed that the global F-theory constraint that there be no more than 24 7-branes is lifted if the space is expanding, even in the simplest case where the expansion is homogeneous and without acceleration.
\footnote{Related spacetimes were considered in \cite{bs1,bs2,bs3}, where
the spatial slices were compact, rather than noncompact, quotients of
two-dimensional hyperbolic space.}

In this note we will use an ansatz similar to \cite{klebanexpand, gruz} to demonstrate that the mild form of SUSY breaking induced by a uniform, constant velocity expansion suffices to allow regions containing a very wide variety of effective string theories to simultaneously co-exist in the same causally connected physical space.  The main examples we consider
are locally flat space, and so (thought of as string solutions) are $\apr$-exact.  However the expanding space has non-trivial topology, and at early times when its cycles are small quantum effects will become important.  We will not attempt to study our solutions at early times; instead, we will focus on their medium and late time behavior (but see \cite{stanford, bs1,bs2,bs3,bs4,bs5}).

Our solutions have implications for the string theory landscape \cite{landscape}, which in turn plays a crucial role in
stringy cosmology.  The application
of landscape ideas to cosmology requires that the large number of
vacua in string theory admit dynamical solutions interpolating among them.
The best known such solutions are Coleman-de Luccia instantons \cite{cdl}
and their string-theoretic cousins, the Euclidean D-brane solutions that
interpolate among flux vacua \cite{teitelboim, bp}.
Some of the most phenomenologically promising compactifications of
string theory are the F theory solutions sketched in \cite{kklt}
and refined in subsequent work.
Dynamical solutions have been found
\cite{freywilliamslippert, kpv} describing transitions from such vacua to
other vacua with different flux numbers, or to noncompact ten-dimensional space
altogether.

At present, very little is understood about cosmological
interpolating solutions
beyond the level of effective field theory.  The description of such
solutions in terms of low-energy dynamics of general relativity coupled to
matter is useful for some calculations, but is limited to
studying transitions in a regime where the
transition generates only small perturbations on the background
spacetime.  As a result, effective field theory may not be capable of
describing transitions in which the solution changes qualitatively, for example
in terms of topology or other gross features.  And yet,
the use of the landscape as a theoretical tool requires the study
of transitions among such qualitatively different vacua.

A related issue is that all realistic string vacua \cite{kklt}
involve orientifolds as an essential ingredient in their
construction.  In perturbation theory, string orientifolds have
no dynamics associated with them.
 Consequently, the known
solutions that interpolate between realistic vacua do not change the number, type
or geometry of orientifolds.  In order to connect
realistic, KKLT-type vacua to the rest of the solution space of
string theory, it must be possible to find interpolating solutions that do change the
features of the orientifolds.
We will exhibit an explicit example of such a solution.

Some controlled examples of interpolating solutions have been constructed
in string theory that change not only the geometry of space, but the
total number of spacetime dimensions and the kind of string theory altogether.
These solutions are not instantons, but
instead involve real-time cosmological
solutions interpolating
from an unstable solution to a stable vacuum or another unstable one
(\cite{h1,h2, hs1,hs2,hs3,hs4,hs5,hs6}).  Though not
realistic as string cosmologies, these solutions
demonstrate that interpolating cosmological solutions among
string vacua exist, and can be controlled and solved exactly even when
they describe changes to the background that are not small perturbations
of a fixed geometry.

The solutions presented here differ from these in several ways, among them that the interpolation is spacelike---the different solutions coexist in regions separated by
spacelike connecting regions.  Specifically we make use of the theory of
\it cobordisms\rm, focusing on the case where the cobordism manifold is hyperbolic, to investigate how the topology of the
compact dimensions can vary continuously from
region to region in the universe.  An interesting fact
is that this mechanism involves only classical geometry, without
direct inputs from string theory.
After developing this framework, we will apply it
to construct a solution interpolating between oriented and unoriented string theories.
In all the cases we consider there are no horizons
in our geometries; the different regions are causally connected and the domain walls separating them are traversible.

\section{Riemann surfaces}

 Begin with the Milne-type metric
\be
\label{metric}
ds^2 = -dt^2 + t^2 dH_n^2 + ds_{m}^2,
\ee
%REMOVED AN H2 SUBSCRIPT FROM DS^2 MKCHANGE
where $dH_n^2$ is the metric on the unit $n$-dimensional hyperbolic space (the maximally symmetric $n$-space with constant negative curvature), $ds_m^2$ is some Ricci-flat $m$-manifold, $t$ is time, and $D=n+m+1$ is the total spacetime dimension.  As can be easily seen this metric is constitutes an exact solution to the vacuum Einstein equations with vanishing cosmological constant,
assuming the internal metric $ds\ll m\sqd$ is such a solution as
well.  In fact, the first two terms in the metric constitute the Milne coordinatization of Minkowski space.  Since the Friedmann-Robertson-Walker scale factor of the expanding part of the metric is $a(t)=t$, the corresponding Hubble length is just $t$, as is the 
proper radius of curvature of the hyperbolic spatial sections.

Hyperbolic $n$-space, being maximally symmetric, has a large isometry group: $SO(n,1)$, with $n(n+1)/2$ generators.
One can construct topologically non-trivial, locally
hyperbolic spaces by identifying $H_n$ under the
action of a discrete subgroup $\Gamma$ of the isometry group.  The result is a space that can have either finite or infinite volume, and, if the subgroup action is properly discontinuous (no fixed points), the space is smooth and free of singularities or defects.

Because the spatial slices remain locally hyperbolic after the identification, the resulting spacetime is still locally flat and therefore a vacuum solution to Einstein's equations and an $\apr$-exact solution in string theory.  However the presence of non-trivial topology means that the full space is no longer globally Minkowski space, and the physics is time-dependent.  At early times, the lengths of the smallest non-contractible cycles of the space become small. The curvature of the
spatial slices becomes large both in string and in Planck units,
and the physics is non-perturbative.  We will focus on later times, where perturbative effects are small and the solution is under full control.  As time passes the cycles grow larger and the solution becomes closer and closer to flat and more and more supersymmetric.

As we will discuss later, for $n>3$ one can consider more general spacetimes where the locally hyperbolic space is replaced by a general Einstein manifold with constant negative curvature.  In that case the full spacetime is a solution at least to lowest order in $\apr$.\footnote{We thank F. Denef for discussions on this point.}

\subsection{Worldwide pants: The expanding triniverse}

To illustrate the possibilities allowed by this class of solutions we begin with a very simple case:   $n=2$, choosing the subgroup $\G$ so that the space $X = \IH\ll 2 / \G$ is a pair of pants: a non-compact Riemann surface with Euler character $-1$, a hyperbolic metric, and a hyperbolic boundary that consists of three disconnected circles (Figure \ref{trin}).  In appendix \ref{appb} we explicitly construct the generators that realize this ``hyperbolic trinion"; $\G$ is a freely acting, finitely generated subgroup of  $PSL(2,\IR)$ in which
every non-identity element is hyperbolic. For simplicity we focus on the case where the trinion has a $\IZ_3$ symmetry that rotates the legs, in which case it turns out to have only a single real modulus preserving the symmetry.    For now we will treat the $m$-dimensional transverse space as 7-dimensional Euclidean space, having in mind weakly coupled critical string theory.

\begin{figure}[b!]
\begin{center}
\includegraphics[width=8cm]{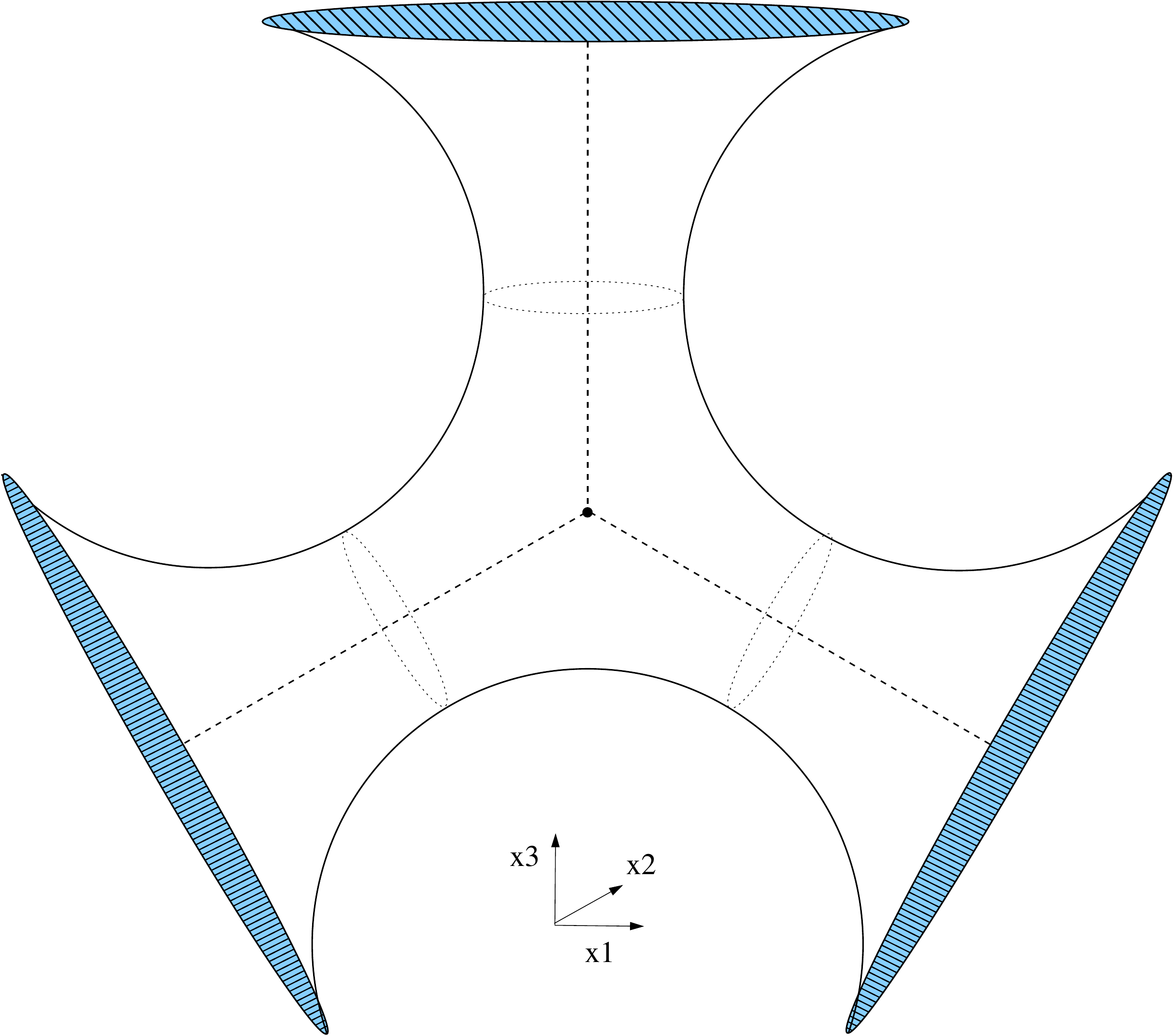}
\end{center}
\caption{\label{trin} \it \small The trinion, or pair of hyperbolic pants.  The dashed circles indicate the three garters---minimum length geodesics wrapping the legs.}
\end{figure}

The 2+1 dimensional part of the spacetime (\ref{metric}) is  a homogeneous but anisotropic
comology.  The spatial slices
are the surfaces $X$, which expand
homogeneously at constant velocity (with scale factor $a(t) = t$).
The metric on $\IH\ll 2$ is the Poincar\'e metric
\bbb
ds\sqd  = {{dx\sqd + dy\sqd}\over{y\sqd}} =
{{dz\cc d\zb}\over{{\rm Im}(z)\sqd}} = {{dr\sqd}\over{r\sqd\cc {\rm sin}\sqd
\th}} +
{{d\th\sqd}\over{{\rm sin}\sqd\th}}\ ,
\een{poincaremetric}
where
\bbb
z = x + i y = r\cc \exp{i\th} \ , \llsk \zb = x - i y = r \cc \exp{- i \th}\ .
\een{threecoords}
The Poincar\'e metric
has Gaussian curvature $-1$, or Ricci scalar curvature $-2$.
The slices have constant negative curvature and
thus the vacuum Einstein equations are satisfied automatically
for our choice of scale factor, for any choice of $\G$.

Prior to taking the quotient by $\G$, the 2+1 dimensional part of the
spacetime is simply
Minkowski space, which means that our cosmological solution
is a quotient of Minkowski space by a discrete subgroup of
the three-dimensional Poincar\'e group acting on $\IR\uu{2,1}$.
This guarantees that the spacetime solves not only the equations of
classical general relativity, but is also a tree-level solution to
string theory to all orders in $\apr$, assuming the transverse $7$-dimensional
factor is also an $\apr$-exact CFT.  This is a somewhat similar
situation to that of \cite{Cornalba:2002fi, sl1, sl2} in that the description of
the spacetime as a quotient of Minkowski space by a subgroup of
the isometry group guarantees its existence as a tree-level solution
to string theory, with corrections coming only from higher-genus
amplitudes.  The loop corrections in our cosmology are more
important than in \cite{sl1,sl2}, since there the subgroup is parabolic and
preserved a lightlike Killing vector, and therefore has vanishing
particle production from the vacuum.  The group $\G$ that corresponds to a trinion is hyperbolic, has
no timelike or lightlike Killing vector, and allows particle production from
the vacuum due to higher-genus effects.  However we will only consider
our solution at tree-level in string theory; as with many FRW cosmologies in
string theory and field theory,
there is a Big Bang singularity at which quantum effects are potentially
important, but any particle production from the initial singularity
is irrelevant at late times, as any density of produced
particles is rapidly redshifted away by the expansion of the universe.

As we will see, with an
appropriate choice of moduli and spin structure, this space can describe a universe in which the physics in different regions
is approximated by a static supersymmetric or non-supersymmetric string theory.   These regions are connected by a Hubble-sized piece of spacetime that functions
qualitatively like a ``domain wall", in the sense that it separates regions with differing effective physics.\footnote{We caution that this region
is quite different from the ``thin-wall" picture of a domain boundary; not only does it have super Hubble-sized thickness, but its effective
dimensionality can be larger than that of either of the two regions that it connects!}
If in addition to modding by $\G$ one performs an orientifold
projection, the space describes a domain wall connecting a region of oriented string theory to an unoriented string theory.

\begin{figure}[b!]
\label{nonagon}
\begin{center}
\includegraphics[width=8cm]{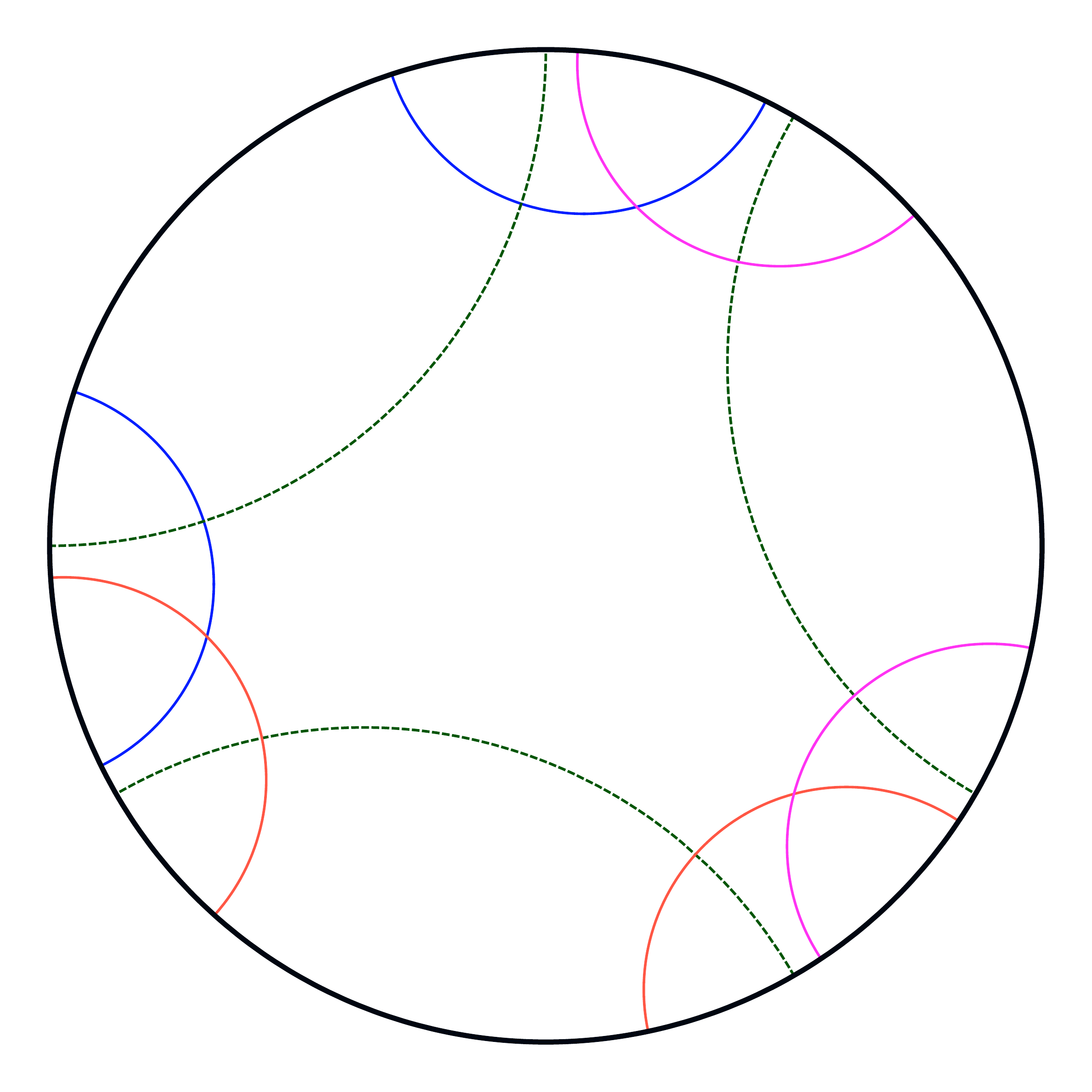}
\end{center}
\caption{\label{trind} \it \small A $\IZ_3$ symmetric trinion as a quotient of the Poincare disk.  Solid curves of the same hue are identified.  The dashed lines are minimal length geodesics; cutting along them would produce a  pair of hyperbolic pants with the skinniest possible cuffs and waist. }
\end{figure}

\begin{figure}[b!]
\begin{center}
\includegraphics[width=8cm]{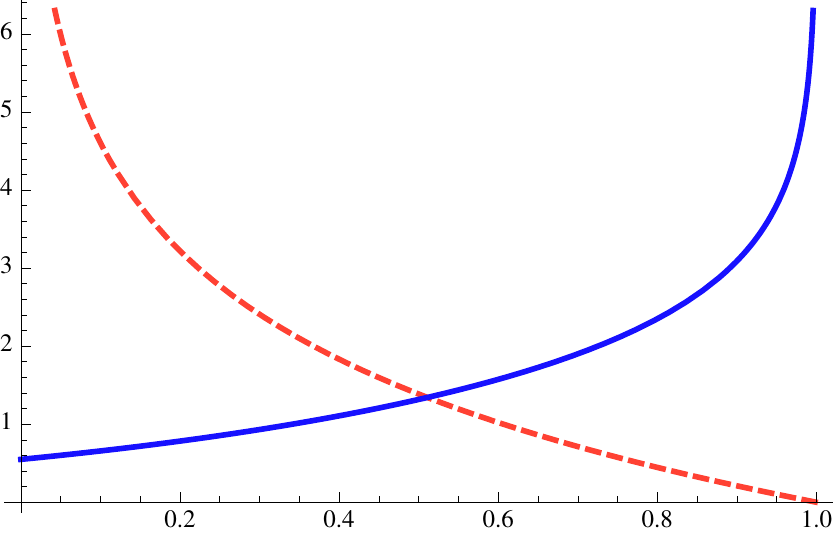}
\end{center}
\caption{\label{lengths} \it \small The solid curve is the geodesic distance from either $Z_3$ fixed point to the minimum length geodesic wrapping a leg (the garter) as a function of the modulus $\gamma$ (see Appendix \ref{appb} for details).  The dashed curve is the geodesic circumference $L$ of the garter, and also the geodesic distance around the crotch (the zipper) between the two $Z_3$ fixed points.  In the limit of interest the garter moves very far from the center, the leg gets very skinny ($L\rightarrow 0$), and the size of the center (measured by the distance between the two $Z_3$ symmetric points) also gets small.}
\end{figure}

%}

\subsection{Garter theory}

Each leg of the hyperbolic trinion (Fig. \ref{trin}) has a geodesic ``garter" where
the 1-cycle wrapping the leg attains its minimum co-moving length $L$.  As we demonstrate in Appendix \ref{appb}, we can choose the modulus of the $\IZ_3$ symmetric trinion to set $L$ to any value from zero to infinity  (Fig. \ref{lengths}).  In particular we can choose $L \lleq 1$ (recall that our hyperbolic space has unit comoving curvature) so that the garter's length is very small in units of the spatial curvature.

The proper length around the garter  in the expanding metric (\ref{metric}) is $a(t)L = tL$, and the proper radius of curvature of the spatial slices is simply $t$.  Hence if we choose the modulus so that $L \lleq 1$, the physical length of the geodesic will always remain small in units of the physical spatial curvature.  However if there is some fixed physical length scale in the dynamics, such as the string length $\sqrt{\apr}$,
the proper radius of the garter will always exceed it at sufficiently late times.  Nevertheless, as we will see small $L$ means that the physics in the vicinity of the garter can be well-described as a compactification on a small circle for very long times and over very long distance scales.

The space in a particular leg
is locally a circle fibered over an interval, with metric
\be \label{coshmet}
ds\ll 2 \sqd = - dt\sqd + t\sqd \lrdd du\sqd + L\sqd \cc {\rm cosh}\sqd u\cc
d\chi \sqd \rrdd\ ,
\ee
 where $\chi$ is identified with $\chi + 1$ ({\it c.f.} Eq. \ref{legb}).
The constant $L$ is a parameter of the solution, expressing the comoving length of the garter in
units of the spatial curvature.  The limit of
interest is $L \<1$.

  For small $u$ the space is an expanding cylinder.   A feature of the small $L$ limit of the modulus is that the region near the minimum length geodesic is many Hubble lengths away from the center of the trinion (Fig. \ref{lengths}).  In that limit the space remains approximately a small circle cross an interval over a very large distance.

 It is illuminating to consider the metric at late times, $t  \simeq t\ll 0 >> 1$, expressing
the comoving length $L$ in terms of the proper radius of the circle, as $L =  2\pi \bfrzero  /t\ll 0$ and exchanging the comoving transverse coordinate $u$ for the proper transverse
coordinate $v \equiv u \cc t\ll 0$.  Then, restoring the factors of the
 speed of light $c$, the metric becomes
\bbb
ds\ll 2 \sqd = - c\sqd \cc dt\sqd + (t\sqd / t\ll 0 \sqd)\cc \lrdd
dv \sqd + 4\pi\sqd \bfrzero \sqd \cc {\rm cosh}\sqd \lrdd {{v }\over{c\cc t\ll 0}}\rrdd
\cc d\chi\sqd
\cc \rrdd\ .
\eee

The garter lies at $v = 0$.  The approximate metric near the garter is
\be
ds_{2}^2 = - c\sqd\cc dt\sqd  + (t\sqd / t\ll 0 \sqd)\cc \lsqq
dv \sqd + 4\pi\sqd \bfrzero \sqd \cc \lrdd 1 + {{v\sqd}\over{2 \cc c\sqd t\ll 0\sqd}}
 \rrdd d\chi \sqd \rsqq   + {\cal O}({{\bfrzero \sqd \cc v \uu 4}\over{t\ll 0\uu 4}})  \cc .
\ee

In the vicinity of the garter, string theory on this space has a spectrum of winding and and Kaluza-Klein modes, with mass-squared given by
  \bbb
  m\sqd = M\ll{10} \sqd + {{\hbar\sqd n\ll{KK}\sqd }\over{c\sqd \bfr\sqd}} + {{\hbar\sqd
  w\sqd \bfr\sqd}\over{ c\sqd \apr\sqd}}\ ,
  \eee
  where $\bfr \equiv {{\bfrzero t}\over{t\ll 0}} \cc {\rm cosh}\lrdd {{ v}\over{ c\cc\t\ll 0}} \rrdd $ is the local radius of the $\chi$ circle,
  $M\ll{10}\sqd$ is the ten-dimensional mass-squared of the string mode,
  and $n\ll{KK}$ and $w$ are the Kaluza-Klein number and winding number of the mode
  around the $\chi$-circle.

  The winding modes feel a confining potential due to the fact that the radius of the compact circle grows quadratically with the coordinate $u$ as one moves away from the garter; conversely, KK modes feel a repulsive quadratic potential.  However the gradient of this potential is weak and $L$-independent, so that the time scale of the instability for KK modes is only Hubble.

To see this, examine the proper acceleration
\bbb
{\rm acceleration} \simeq {{c\sqd}\over m} {{dm}\over{dv}}
\eee
for a KK mode of a massless
10-dimensional field; the mass is $m_{{KK}}= {{\hbar \cc n\ll {KK} }\over{c\cc \bfr}}$, so
the acceleration at position $v$ will be
\be
{\rm acceleration} \simeq  v /t\ll 0 ^2 \ ,
\ee
to lowest order in $v$, with corrections of order $ {{v \uu 3}\over{c\sqd\cc
t\ll 0\uu 4}}$.  For $v$ in the vicinity of the garter $v \<  c\cc H\uu{-1}$, a KK mode is approximately inertial over time scales comparable to the age of the universe.

  The acceleration of a wound string has the same magnitude but the opposite sign.  The magnitude of the proper acceleration of either a KK or wound string mode within a Hubble length of the garter is at most the Hubble acceleration.  In the limit of interest to us ($H \< m_{s}$), these accelerations are very slow.

 The regions near distinct garters never go out of causal contact with each other; that is, there are no horizons in this cosmology.  The scale factor $a(t)=t$ is the marginal case of FRW expansion with neither acceleration nor deceleration.  The FRW time it takes light to propagate between two comoving points (such as two of the garters) is proportional to the start time and exponential in the comoving distance between the points (see Appendix \ref{causgeo} for details).

 A string mode created near one of the garters could be fired towards the crotch, resulting in (potentially painful) scattering and transmission amplitudes off the ``domain wall" separating the legs.  Amusingly, the worldsheet corresponding to this process would itself be a trinion that clothes the spacetime trinion.  It would be interesting to study such processes, particularly in cases where the effective theories near the garters are very different (see below), but we will not pursue this further here.

\subsection{Spin structure and dynamical stability}

Let us now comment on the issue of the dynamical stability of our solution.  Because of
the expansion of the universe the question of ``stability" must be treated with some care;
the background is explicitly time-dependent.  However if we focus on
the dynamics in a garter region with $\bfrzero \< c\cc H\uu{-1}$, we can ask about
dynamical stability on time scales of order $  \bfrzero / c$, in which case the
background can be treated as if it were static.
As we have seen, in this limit each garter region becomes the product of a line (parametrized by $v$)
 and a circle (parametrized by $\chi$).  So we can
understand the dynamical stability of each region in terms of the limiting static
compactification on $S\uu 1$.

In this limit, the dynamical stability of the compactification is determined by
the radius $\bfrzero$ of the circle, the amount of SUSY in the underlying
theory, and the boundary conditions for the gravitini along the circle in a given
garter region.  If the underlying theory has a bulk tachyon---such as in
type 0 or bosonic string theory---every garter region will be nonsupersymmetric.

If the underlying theory has ${\cal N} = 1$ or ${\cal N} = 2$ supersymmetry, then
a garter region may be stable or unstable, depending on the size of the
$\chi$-circle and the boundary conditions on the gravitini.
If one or both gravitini are periodic around the $\chi$-circle,
the garter region will have approximate ${\cal N} = 1$ or ${\cal N} = 2$
supersymmetry, respectively.  The SUSY in this case is broken
 only by the overall expansion of the universe, so the scale of
 SUSY breaking $\hbar \cc H$ is the same scale as that
 of the time-variation of the background itself.

 If the underlying theory has ${\cal N} = 1$ or ${\cal N} = 2$ supersymmetry but
 both gravitini are antiperiodic around a given garter, the theory may still be
 tachyon-free, depending whether or not
 the radius $\bfrzero$ of the garter is large compared to the
 fundamental scale of the underlying gravity theory.

 In the case of 11-dimensional
 ${\cal M}$-theory, a garter region with antiperiodic gravitini should be
 tachyon-free so long as the garter radius $\bfrzero$ is large compared to the only
 scale in the system, the $11$-dimensional Planck scale $L\ll{11}$.
 If $\bfrzero$ is
 much less then $L\ll{11}$ we expect that the effective 10-dimensional theory is described by
 type 0A string theory at weak coupling; according to
 the standard worldsheet analysis of this system, there is a tachyon with
  a rapid decay constant, $\G\ll{\rm tachyon}  =  c\cc \sqrt{{2\over{\apr}}}$
  (which in terms of 11-dimensional data\cite{edm} is proportional to $c\cc L\ll{11}\uu{-{3\over 2}}
  \cc \bfrzero \uu{+\hh}$).

 In the case of 10-dimensional type IIA/B string theory with antiperiodic boundary
 conditions for all the fermions around the garter, the dynamical stability of the region
  can be analyzed systematically when the coupling is weak
  with string perturbation theory.  Using the results of \cite{rohm, bergman} we see
  that the lowest winding string mode has mass $m\sqd = - {2\over{\apr}} + {{\bfrzero\sqd}
  \over{\apr\sqd}}$ at string tree level. So, for $\bfrzero > \sqrt{2 \apr}$, the compactification is tachyon-free at time $t\ll 0$; for $\bfrzero < \sqrt{2\apr}$ there is
  a vacuum instability with a decay rate of order $c / \sqrt{\apr}$.

  Even the tachyon-free regimes of 11-dimensional M-theory and 10-dimensional string
  theory with antiperiodic gravitini are not truly stable.  There are nonperturbative gravitational instanton solutions \cite{wittenbubble} that allow a nonsupersymmetric leg
  to pinch off, disconnecting the asymptotic region from the 3-leg junction (although
  the action for such an instanton increases with the size $\bfrzero$ of the circle,
  so the integrated decay probability over all time will be finite).
  Furthermore, at one loop one expects that
  the mismatch of boundary conditions for bosons and fermions will lead to nonvanishing
  Casimir energies, generating a potential for the modulus $\bfrzero$ that scales as a
  power of $\bfrzero$ when $\bfrzero$ is large compared to the fundamental scale.
  In order to arrange truly stable dynamics, it is necessary for
  at least one gravitino to be periodic in each garter region.

   How much freedom does one have to achieve such a configuration?  For each
   spinor's worth of local supersymmetry, one has the freedom to choose a spin structure determining the boundary conditions for fermions around the garter.  Since in this example there are three garters (one on each leg of the trinion), naively there are three independent
choices to make to determine the spin structure.  In fact this is not quite correct:  only two of the three cycles are homologically independent, so the the boundary conditions
for fermions on the three cycles must satisfy a single relation---that the product of periodicities around all three garters multiply
 to -1 for each gravitino.   The consequence is that
either one or all three of the boundary conditions for a given gravitino field must be anti-periodic, as can easily be seen by considering the thrice-punctured sphere (the cycle that encircles all three punctures is contractible, and therefore fermions going around it must be anti-periodic).
So it would seem that at least one of the three legs must always have ``large" amount of
supersymmetry breaking, characterized by the SUSY breaking scale $m\ll{KK} = {1\over{\bfrzero}}$.

Such is the situation when one considers a theory with ${\cal N} = 1$
supersymmetry (such as heterotic or type $I$ string theory in 10 dimensions or
11-dimensional $M$-theory).  These theories do not admit solutions with supersymmetric
boundary conditions in each leg of a three-universe junction.  (We shall see later that
this constraint is an artifact of a geometry with three asymptotic regions
rather than another number, and can be evaded easily by adding extra legs to the
Riemann surface.)

 When one considers theories with ${\cal N} = 2$
supersymmetry such as type IIA or IIB string theory,
there are two independent 10D gravitini and two independent
spinor fields' worth of local SUSY,
 and therefore twice as many choices of spin structure on the trinion, one for each
  gravitino.  Each spin structure must separately
 satisfy the consistency condition, with antiperiodic boundary conditions around at
 least one leg; however we have the freedom to choose the two gravitini to be
 antiperiodic around \it different \rm legs, so that each leg has at least one periodic
 gravitino and therefore at least one dynamical supersymmetry preserved up to
 Hubble-scale effects.

 For instance, we may choose the first set of gravitini to be periodic around garters A and
 B and antiperiodic around garter C, and the second set of gravitini to be periodic
 around garters B and C, and antiperiodic around garter A.  The effective dynamics
  near garter $B$ has approximate ${\cal N} = 2$ supersymmetry in $9$ dimensions, while
  the effective dynamics near garters $A$ and $C$ respect only ${\cal N} = 1$ supersymmetry
  in $9$ dimensions.  The latter is described by a chiral Scherk-Schwarz compactification
  of the type described in \cite{chiralgutperle, chiralhellerman}.  Of course, the
  surviving SUSY in any region is merely approximate, being broken cosmologically
  by the overall expansion of the universe.  But in the regime $L\< 1$ the
  distinction between a supersymmetric and nonsupersymmetric leg of the trinion is
  very sharp.  In a supersymmetric leg, the SUSY breaking is Hubble-scale, with
  mass splittings at most of order $\Delta m \simeq H$ in each multiplet, whereas in the
  nonsupersymmetric legs the splittings would be of order $\Delta m \simeq {1\over{\bfrzero}}$.

\begin{figure}[b!]
\begin{center}
\includegraphics[width=8cm]{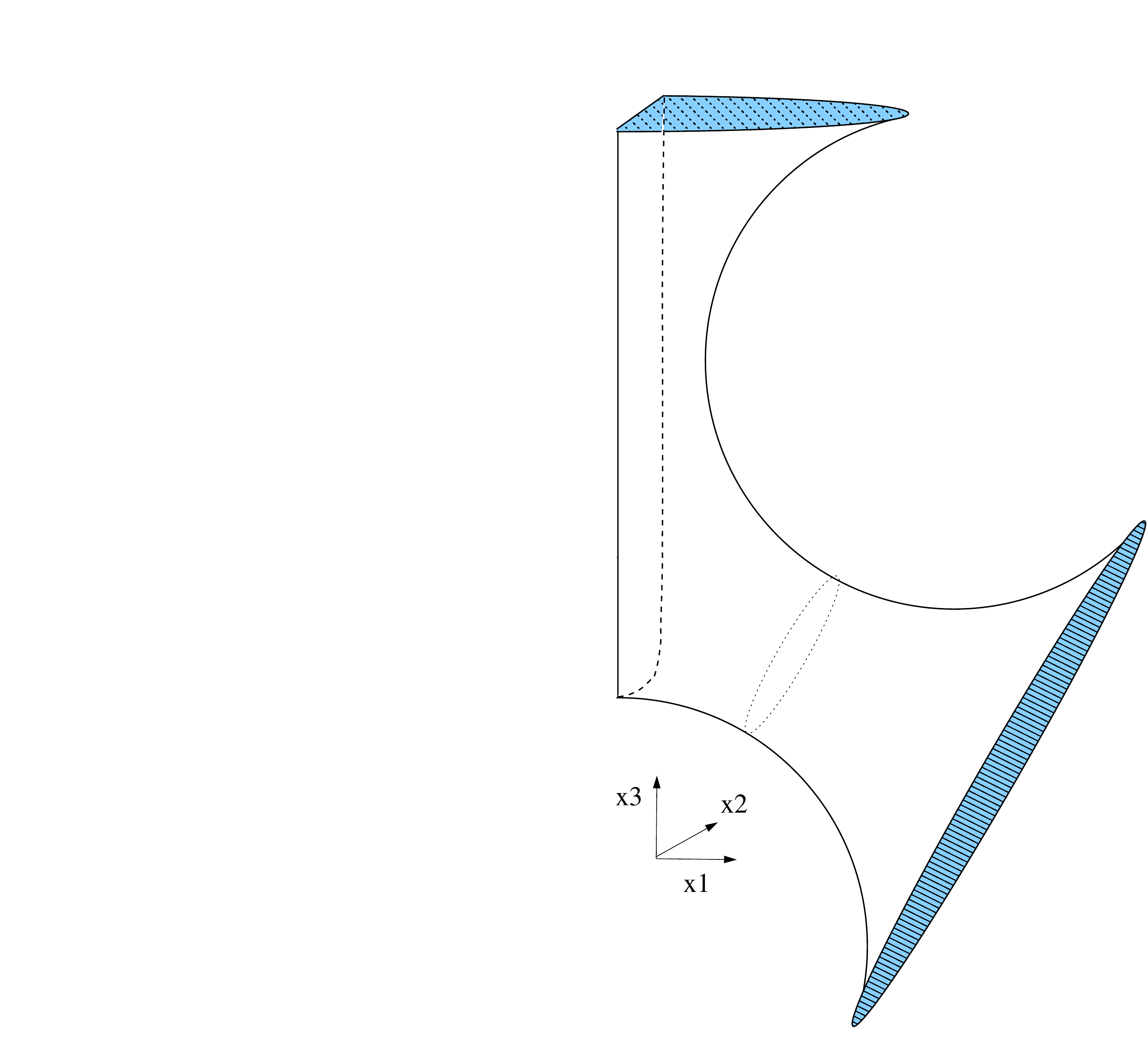}
\end{center}
\caption{\label{triZ2} \it \small The orientitrinifold: the quotient of 
a three-legged Riemann surface by a $\IZ\ll 2$ symmetry reflecting one pair
of legs into one another, and the third leg into itself.}
\end{figure}

%}

\subsection{Orientifolds}

If the trinion has a $\IZ_2$ symmetry that reflects one leg onto another, one can perform an orbifold or orientifold projection that maps the trinion into a space that connects a two asymptotic regions with different orientability properties for
 string worldsheets.  For definiteness, arrange the trinion so that its legs lie in
 the $x\ll 2 - x\ll 3$ plane, with one leg pointing along the positive $x\ll 3$
 axis, and the other two arranged to point at equal acute angles to the negative
 $x\ll 2$ axis, as in Figure \ref{trin}.  There is a discrete symmetry ${\bf R}\ll 2$ that
 leaves $x\ll 1$ and $x\ll 3$ unchanged, and reflects the $x\ll 2$ coordinate.  For
 the trinion embedded in Euclidean three space as in Figure \ref{trin}, the reflection
 ${\bf R}\ll 2$ acts as an antiholomorphic involution on the trinion, with a
 fixed locus of real dimension 1.  The reflection ${\bf R}\ll 2$ reflects legs A and
 C into one another, and reflects half of leg B into the other half.  The fixed
 locus consists of a hairpin-shaped curve travelling down one side of leg B and up the other side -- in pair-of-pants terminology, thinking of leg B as the waist of the pair
 of pants, the fixed locus coincides with the zipper, but extended infinitely as a geodesic along the front and back of the waist.

 Now let us orientifold the pair of pants by the action of ${\bf R}\ll 2$.  Since the
 fixed locus has real codimension 1, this restricts us to type IIA string theory (unless
 we combine the action of ${\bf R}\ll 2$ with some other action on the seven coordinates
 not involved in the cosmological solution).  Since
 we are considering an orientifold, rather than an orbifold action, there are no consistency
 conditions on the action of ${\bf R}\ll 2$ at string tree level other than the
 condition that the real codimension of the fixed locus be odd.  At higher order in the
 string coupling, the tension and Ramond-Ramond charge of the O8-plane generate
 tadpoles for the dilaton, metric and Ramond-Ramond 10-form flux.  To cancel the tadpole
 we can place 16 D8-branes along the locus of the O8-plane.
 
 The orientifolded spacetime, then, is given by one half of the original spacetime---a
 fundamental region of the action of ${\bf R}\ll 2$, with O8-plane boundaries and with
 coincident D8-branes as desired.  Leg B (pointing along the positive $x\ll 3$-axis) is
 reflected into itself by the action of ${\bf R}\ll 2$, and so it is sliced in half by the
 orientifolding, with two O8-plane boundaries where leg B intersects the plane
 $x\ll 2 = 0$.  Legs A and C are symmetrically
 arranged on either side of $x\ll 2 = 0$ and do not intersect it, so the reflection ${\bf R}\ll 2$ reflects legs A and C
 into one another.  Therefore, the geometry of the spacetime after orientifolding has
 two asymptotic regions: One asymptotic region points at an acute angle to the negative $x\ll 3$ axis with the geometry of a circle fibered over a ray, and the other asymptotic region
 lies along the positive $x\ll 3$ axis, with the geometry of an interval, with O8-plane
 boundaries, fibered along a ray.  This geometry is depicted in
 Figure \ref{triZ2}.

\subsection{More general surfaces}\label{moregen}

\heading{Riemann surfaces with more than three legs}

So far we have focused on the specific example of the hyperbolic trinion.  Our
construction extends equally well to Riemann surfaces obtained as quotients of
$\IH\ll 2 $ by more complicated subgroups of $SL(2,\IR)$.
The closest generalization would be to free subgroups $\G$ of $SL(2,\IR)$ generated by
$b-1$ elements, with every non-identity element of $\G$ being hyperbolic and
chosen such that the quotient $X\equiv \IH\ll 2 / \G$ is a sphere with $b$ or more holes ($b \geq 4$).  There are $3b-6$ real moduli describing the embedding of $\G$ in $SL(2,\IR)$:
for each of the $b-1$ independent generators, there is a choice of element of
$SL(2,\IR)$, with each specified by three parameters.  The freedom to conjugate the elements of the generating set by a  general $SL(2,\IR)$ matrix subtracts
three parameters, leaving $3b-6$ real parameters specifying the identifications modulo
coordinate transformations.  We can equally well describe the geometry as the unique
negatively curved metric on a Riemann surface described as a sphere with $b$ circular boundaries.  The structure of the complex manifold in this description is specified by
the locations of $b$ points, and the radii of the $b$ holes on the sphere, modulo
an overall $SL(2,\IC)$ transformation, for a total of $2b+b - 6 = 3b-6$ real parameters.
We can of course impose additional discrete symmetries to cut down the number of moduli,
and we can orbifold or orientifold by these symmetries if we choose.

Any such Riemann surface $X$ generates an exact
solution of the vacuum Einstein equations and an $\apr$-exact solution of string theory, as
was the case for the trinionic spatial slices we discussed earlier.  The metric
\bbb
ds\sqd\ll{\rm FRW} = - dt\sqd + t\sqd \cc ds\ll X\sqd
\eee
is an orbifold of $2+1$-dimensional Minkowski space by a discrete group $\G$, and thus gives
rise to a direct construction of an exactly conformal 2-dimensional worldsheet field theory
describing string propagation.

  The extension to $b$-universe junctions with $b > 3$ makes it clear that the selection rules on boundary conditions for gravitini are very weak, and can in effect be
   evaded by adding additional asymptotic regions.  The selection rule on spinor periodicities is that, if $\s\ll a$ is the periodicity of a gravitino around the garter
  in the $a\uu{\underline{\rm {th}}}$ leg (with $-1$ for antiperiodic fermions and
   $+1$ for periodic fermions), then the set of periodicities must satisfy
   \bbb
   \prod\ll{a = 1}\uu b \cc (- \s\ll a) = +1\ .
   \eee
   In particular, for $b$ even it is possible to choose all garter regions to be supersymmetric.

\heading{Solutions interpolating between type IIA and $E\ll 8 \times E\ll 8$ heterotic
string theory}

Combined with the freedom to take $\IZ\ll 2$ quotients with codimension-1 fixed planes, the case $b=4$ of the geometry discussed above
gives us a construction of interpolating solutions between
supersymmetric regions of different types in theories with ${\cal N} = 1$ supersymmetry,
in particular 11-dimensional M-theory.

\begin{figure}[b!]
\begin{center}
\includegraphics[width=12cm]{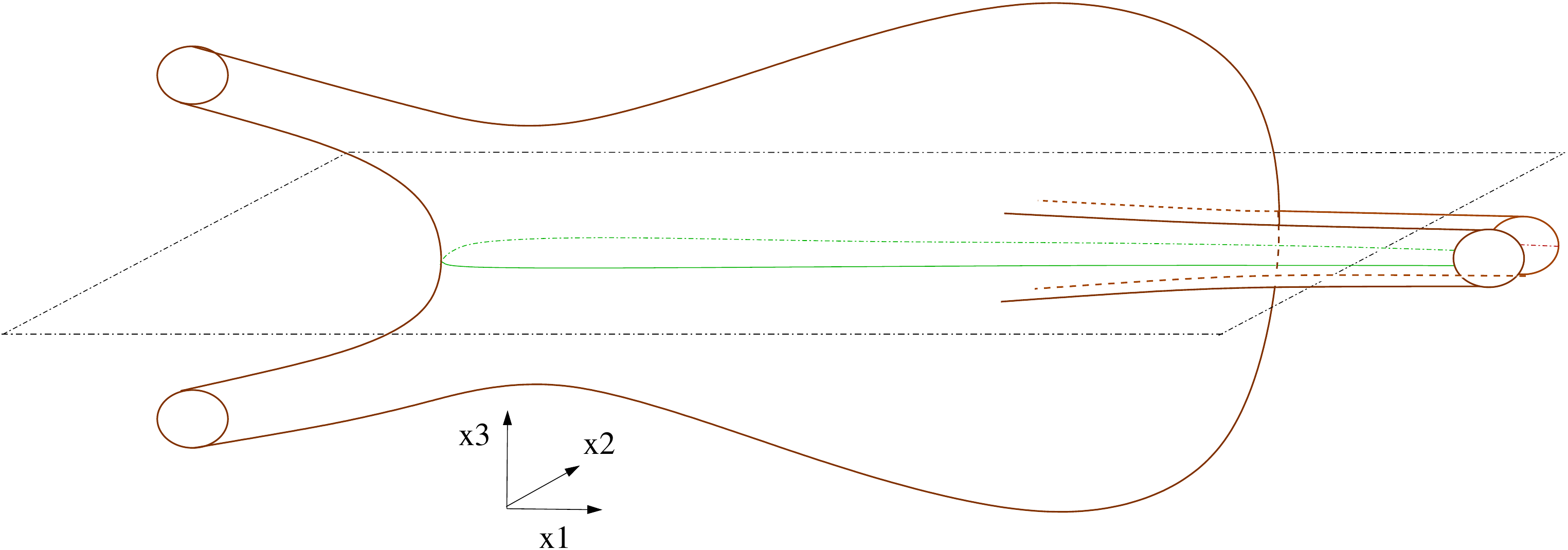}
\end{center}
\caption{\label{quad} \it \small A Riemann surface with four hyperbolic legs,
has the geometry of a whale with two $\IZ\ll 2$ symmetries.
Each $\IZ\ll 2$ symmetry factor exchanges one pair of legs and
reflects each of the other pair into itself.  A quotient by one of the $\IZ\ll 2$ symmetries has a single
leg with circular cross-section and two legs with interval cross sections. }
\end{figure}

We begin with a Riemann surface with four legs (depicted in Figure \ref{quad}),
and a set of reflection symmetries realized in a particular way. To
 describe the symmetries conveniently we embed the surface in
three-dimensional Euclidean space, with axes labelled $x\ll{1,2,3}$.  The two legs
pointing in the negative $x\ll 1$-direction are arranged to lie in the $x\ll 1 - x\ll 3$
plane, while the two legs pointing in the positive $x\ll 1$ direction are arranged to
lie in the $x\ll 1 - x\ll 2$ plane.  One reflection symmetry (${\bf R}\ll 2$) fixes
$x\ll 1$ and $x\ll 3$, and reflects the $x\ll 2$ coordinate, while the other
reflection symmetry (${\bf R}\ll 3$)  fixes
$x\ll 1$ and $x\ll 2$, and reflects the $x\ll 3$ coordinate.  Both symmetries act
as antiholomorphic involutions on the Riemann surface, with fixed loci of real codimension 1. The reflection ${\bf R}\ll 3$ reflects the first pair of legs into one another,
and reflects each of the second pair of legs into itself, while the reflection
${\bf R}\ll 2$ reflects the second pair of legs into one another, and reflects each
of the first pair of legs into itself.

%}

 Having defined this Riemann surface and the associated cosmological solution of general relativity/string/M-theory, we can orbifold the solution by one of its two reflection symmetries, say ${\bf R}\ll 3$.  The resulting quotient has a single
 cylindrical leg with closed circular cross-sections in the asymptotic region at negative $x\ll 1$,
 and two regions which are topologically intervals fibered over a ray in the asymptotic
 region at positive $x\ll 1$.  This geometry is a consistent exact
 solution of any gravity theory that
 allows orbifolding by a reflection with a real codimension-1 fixed locus, adding
 branes for tadpole cancellation as necessary.  Theories
 allowing such orbifolding/orientifolding
 include type IIA string theory and 11-dimensional M-theory.

  The case of M-theory on such an orbifold is particularly interesting, as it is known
  \cite{hw1, hw2} that codimension-1 orbifold fixed planes in M-theory support propagating
  $E\ll 8$ gauge symmetries.  M-theory compactified on an interval bounded by such fixed
  loci is described by weakly coupled $E\ll 8 \times E\ll 8 $ heterotic string theory in
  the limit where the length $\pi \bfrzero$ of the interval at its minimum is much smaller
  than the 11-dimensional Planck length $L\ll {11}$, assuming the boundary conditions on
  the gravitini are supersymmetric, as we have arranged them to be in this case.

  This geometry, and its realization as the spatial section of an FRW metric, gives an explicit construction of a controlled solution of
  M theory that realizes dramatically different phases of the theory in
   different regions of the universe; one asymptotic region is described by
   the dynamics of type IIA string theory, where the other two asymptotic regions are described by the dynamics of heterotic string theory.  The size of the interval and that of the circle grow exponentially away from
   the garter region.  However the exponential growth as a function of distance from the
   garter is slow, with variation only on Hubble distances.  Thus if we tune all
   the garter lengths to be much smaller than the 11D Planck scale, then at late times
   we have macroscopically sized regions of different superstring theories realized in
   different regions of the same universe, all causally connected to one another.  There is
   no contradiction from the point of view of string theory, as there is always a
   region in the middle that is strongly coupled from the point of view of either set of string
   degrees of freedom.  Nonetheless the solution at late times is protected from quantum corrections due to the hierarchy between the Hubble scale and the Planck scale in the regions
   where the geometry is large, and the residual supersymmetry in the regions where the
   geometry is small.

   \heading{Cusps and parabolic quotients}
% REWROTE THIS SECTION A BIT.  I CHANGED THE LAST SENTENCE FROM LEADING ORDER IN ALPHA PRIME TO EXACT IN ALPHA PRIME.  MAKE SURE YOU AGREE.  ALSO MAYBE STILL NOT AS CLEAR AS IT COULD BE  MKCHANGE

  So far we have studied the physics near the garter of one of the legs of a multi-legged, negatively curved Riemann surface.  It is also illuminating to expand around one of the legs away from the minimum at the garter.   Expanding $u$ around $u\ll 0 > 0$ as $u=u_0 + \epsilon$ with $u_0 \ggg 1$, $\cosh(u)\approx e^{u_0+\epsilon}/2$ and the metric (\ref{coshmet}) becomes
\bbb \label{cusp}
ds\ll 2 \sqd  \approx -dt^2 + t^2 \left[ d\epsilon^2 + \left( {L e^{u_0} / 2} \right)^2 \left(1 + 2 \epsilon + {\cal O}(\epsilon^2)\right) \right]d\chi^2.
\eee
The space near the garter $u=0$ is symmetric under reflections in $u$, which results in a quadratic potential for winding and KK modes.  Instead here there is a linear term in the potential and no symmetric point---evaluating the proper acceleration of a KK or winding mode, one finds simply
\bbb
{d \ln(m_{KK})/( t d\epsilon)} \approx 1/t = H.
\eee
So even away from the garter the variation in the leg radius can be neglected, at least for times and distances short compared to the Hubble scale.  When $L e^{u_0}$ is sufficiently small, at time $t_0$ the physics within a Hubble radius of $u=u_0$ will be well approximated by the KK reduction of a circle of physical size  $t_0 L e^{u_0}$.

As we will discuss briefly below, hyperbolic manifolds generically develop regions known as cusps in
singular limits of their moduli spaces.  These ``thin" regions have metric
\bbb
ds_n^2 \approx du^2 + e^{2 u} ds_{n-1}^2.
\eee
If the cusp cross-section $ ds_{n-1}^2$ is flat (for example a torus), this is locally hyperbolic space.  

Expanding as before around some point $u=u_0$, we have just seen that if $t_0 L_i e^{u_0}$ is sufficiently small (where $L_i$ is the comoving length of some cycle in $ ds_{n-1}^2$) we can reduce on that cycle.  This provides a very general method for constructing $\apr$ exact interpolating string solutions in higher dimensions.

\section{Higher dimensions}

Our ansatz \eqref{metric} was
\be
ds^2 = -dt^2 + t^2 ds_n^2 + ds_{m}^2,
\ee
where so far we have considered cases in which $ds_n^2$ is a Riemann surface.  More generally if  $ds_n^2$ is a locally hyperbolic (constant negative sectional curvature) manifold, and $ds_m^2$ is $m$-dimensional Euclidean space, an orbifold of it, a Calabi-Yau, a product of those, or some other exactly conformal sigma model, this metric is an $\alpha'$-exact solution to string theory (and a solution to supergravity with no higher curvature corrections to the $n+1$-dimensional factor of the geometry.)

Instead of $\apr$-exactness we could require merely that the metric solve the low energy equations of string or M-theory, {\it i.e.} Einstein's equations plus the equations of motion for the other massless fields.  In that case the $n$-dimensional space $ds_n^2$ need not be hyperbolic.  Instead, it should satisfy
\be
R_n^{ij} = -(n-1) g_n^{ij},
\ee
where $i,j \in 1,..,n$ run over the indices of $ds_n^2$; {\it i.e.} it should be an Einstein manifold with Ricci scalar $R=-n(n-1) < 0$.  If in addition $ds_m^2$ is Ricci flat, one can easily verify that the full spacetime satisfies  $R_{\mu \nu} = 0$, where $\mu$ and $\nu$ run over all coordinates. In dimensions $n=2$ and 3 Einstein with negative curvature is equivalent to locally hyperbolic (due to the absence of gravity waves), but for $n>3$ it is a weaker condition.

\subsection{Cobordisms and their significance}

By definition, the topology of a smooth manifold is exactly the collection of
properties of the manifold that cannot be altered by deforming it in any
smooth way.  Therefore it may seem counter-intuitive that it should be
possible to ``interpolate smoothly" between two smooth manifolds of different
topology, at least when classical notions of geometry and topology are in force.  And yet it is indeed possible to do so.

By a smooth interpolation between two $n$-manifolds $X\ll{1,2}$ we mean a
smooth $(n+1)$-manifold $Y$ such that the boundary of $Y$ is equal to the
disconnected sum of $X\ll 1 $ and $X\ll 2$, with the two components understood
as having opposite orientation.  There are many intuitive and
easy-to-visualize examples in the case $n=1$.  For instance, if $X\ll 1$ is a
circle and $X\ll 2$ is the empty set $\emptyset$, then a cobordism $Y$ between
$X\ll 1$ and $X\ll 2$ would be a smooth manifold $Y$ whose boundary is just the circle
$X\ll 1$ (such interpolating manifolds are quite familiar to physicists working on perturbative string theory).

Suppose we wish to find a solution in our class that interpolates between two string compactifications.  In order for such a solution to exist, the two compactification manifolds must be cobordant.  Moreover, the bordism manifold must admit a hyperbolic or negatively curved Einstein metric.  In the following section we will briefly discuss under what conditions this is possible.

\subsection{$n=2$}

By the uniformization theorem, all Riemann surfaces with Euler character $\chi < 0$ admit a hyperbolic metric.  All 1-manifolds are cobordant to each other (although if spin structure is included, there is a $Z_2$ selection ruling requiring that the number of boundary components with period fermions must equal mod 2).

The trinion solution discussed above is an explicit example of such a hyperbolic cobordism, in that case between $S^1$ and $S^1\oplus S^1$.  The orientifolded trinion interpolates between and interval and an $S^1$.

\subsection{$n=3$}

While all negatively curved Einstein 3-manifolds are hyperbolic, not all 3-manifolds admit such a metric, and those that do have a restricted structure.  In fact the space of all 3-manifolds has been fully characterized through geometrization.  All compact 3-manifolds can be decomposed into pieces of eight types: hyperbolic, spherical, Euclidean, and five other possibilities.  In the case of interest for us---hyperbolic---much is known about the possible topologies and metric structures.  For further details, see {\it e.g.} \cite{lectures}.

Because we are interested in regions of the hyperbolic 3-manifold where some cycles shrink, we can make use of the so-called ``thick-thin" decomposition.  This is a technique for dividing the manifold into a region where some cycle(s) get smaller than some universal $n$-dependent length (the ``thin" part), and the rest of the manifold  where all cycle(s) are large (the ``thick" part).

The utility of this decomposition arises from the Margulis lemma, which restricted to $H_2$ and $H_3$ can be stated as follows:
 \begin{quote} There exists a number $\epsilon_n > 0$ such that if $\Gamma$ is a discrete group of isometries of
$H_n$ generated by $\gamma_1 , \gamma_2 , ..., \gamma_k$ and there exists $p \in H_n$ such that $d(p, \gamma_i p) \leq \epsilon_n$, then $\Gamma$ is Abelian.\end{quote}
Here $n=2,3$, and $d(p_1, p_2)$ is the geodesic distance between the points $p_1$ and $p_2$, and the $\epsilon_{n}$ are ${\cal O}(1)$ constants depending only on the dimension $n$.

In the case of real surfaces $n=2$,  Abelian $\Gamma$ means $\Gamma$ is generated by a single generator (since commuting elements of a Fuchsian group must share fixed points).  For a smooth space the generator is either hyperbolic or parabolic.  Therefore the ``thin" part of any hyperbolic surface can be one of only two possibilities:  either a garter (a circle cross an interval with ``$\cosh$" warp factor) or a cusp (a circle cross an interval with decreasing exponential warp factor), where for the former the generator is hyperbolic, and for the latter, parabolic.   The trinion has three thin regions of the first type (the three garters discussed above) connected by and to thick regions (the central junction and the hyperbolic trumpets that go off to the boundary---Figure \ref{trin}).

The case of $n=3$ is somewhat more complex.  The closest analog of the $n=2$ garter is a region of the space around a short closed geodesic cycle  that is the axis of some hyperbolic generator in $\Gamma$.\footnote{A generalization of this is where the generator that translates along $\chi$ is loxodromic (trace squared not in $[0,4]$) rather than simply hyperbolic.  In that case it generates a translation in $\chi$ times a rotation in the angle $\theta$.}  The space can be parametrized locally by distance along the geodesic cycle $\chi \simeq \chi+1$, an angle around the cycle $\theta \simeq \theta+2\pi$, and a radial distance $u$:
\be \label{3cusp}
ds_3^2 \sim du^2 + \sinh^{2}(u)~d\theta^2 + \cosh^2(u)~L^2 d\chi^2,
\ee
where $L<\epsilon_3$ is the distance around the cycle at its shortest point $u=0$ (the garter).  As the radial distance $u$ increases, the length of the closed cycle grows as $\cosh u$.
This type of thin region is compact; its boundary is a flat 2-torus.\footnote{ If the metric remains \eqref{3cusp} over the entire manifold, these spaces are familiar to physicists as the euclidean BTZ black hole \cite{Banados:1992wn}, while the general loxodromic case is related to BTZ holes with angular momentum.  The thin part is simply the tip of the Euclidean cigar geometry cross a circle (the horizon of the BTZ black hole).}

The other type of thin part involves quotients by one or two parabolic elements.  The metric on the thin part is
\be \label{cuspmet}
dz^2 + e^{-2 z}(dx^2 + dy^2)
\ee
where either one or both of $x$ and $y$ are identified under translations: $x \simeq x + L_x$, $y \simeq y + L_y$.   These are cusps, where the cross-section of the cusp is either a cylinder ($L_x$ finite, $L_y$ infinite) or a torus (both $L_x, L_y$ finite).  In the latter case the thin part is non-compact but of finite volume.

%ADDED NEXT FEW SENTENCES PLUS OTHER CHANGES IN THIS SUBSECTION MKCHANGE
The thin regions of hyperbolic type have topology $D^2 \times S^1$, where the radius of the $S^1$ is minimum at the center of the disk---the garter---and increases (gradually, given an appropriate choice of moduli) as one moves radially out.   The region within a Hubble length of the garter  is two dimensional flat space cross a short closed cycle, and so one can reduce on the cycle and obtain a theory in two flat dimensions (parametrized by $u$ and $\theta$ in (\ref{3cusp})), plus the transverse $m$ dimensions of (\ref{metric}).

On the other hand parabolic thin regions are 2-torii cross a semi-infinite interval, where the cycles of the torus can be made arbitrarily small by moving out along the cusp and have approximately constant size within a Hubble patch.  Reducing on both cycles of the torus would give a theory in one dimension (parametrized by $z$ in (\ref{cuspmet})) plus the transverse $m$ dimensions.  Hence a 3D hyperbolic manifold possessing thin regions of each type would connect effective theories in different numbers of dimensions and with different topology for the compact dimensions. 
%COULD INSTEAD SAY - IN WHICH THE EFFECTIVE COMPACTIFICATION MANIFOLD HAS DIFFERENT NUMBERS OF DIMENSIONS OR SOMETHING LIKE THAT MKCHANGE

\subsection{$n>3$}

For $n>3$ we can restrict to locally hyperbolic manifolds if we would like to find $\alpha'$-exact string solutions, or allow for general Einstein manifolds with negative curvature if we only require that the solutions be valid to lowest order in $\alpha'$.

\paragraph{Locally hyperbolic n-manifolds:}  The thick-thin classification above is effective for $n>3$, but there are more possibilities for the thin parts.  Roughly speaking they are either the obvious generalization of the garter case discussed above for $n=3$ (the space near the garter is Euclidean $n-1$ space $E_{n-1}$ cross a small circle, with the size of the circle growing like cosh as one moves away from the origin of the Euclidean space), or cusps with cross-sections that are orbifolds of $E_{n-1}$.  Those include $(n-1)$-torii, and hyperelliptic surfaces and their higher and odd-dimensional relatives.

\paragraph{Negatively curved Einstein manifolds:}  In the case $n=4$ there are some mild conditions on the topology of Einstein manifolds with negative curvature \cite{Besse}.  However for $n>4$ there are no known restrictions: it may be that all compact $n$-manifolds with $n>4$ admit negatively curved Einstein metrics.  If so, we can connect nearly {\it any} collection of $n-1$-manifolds using our expanding solution.  Of course (since this is a topological statement) one is not guaranteed that the geometry of the $n-1$ manifolds can be adjusted as desired.  But given the considerable freedom available, it seems likely that any or nearly any string compactification on (for example) an $n-1=6$ manifold can be approximately realized locally and connected to any other such compactification by a negatively curved Einstein 7-manifold.  In fact it seems plausible that the entire vacuum landscape of string and M-theory can be connected this way.

%COMMENTED THIS OUT MKCHANGE

%One may even be able to realize the famous ``M-theory spiky blob" as a real expanding space-time, where the space is the blob, expanding homogeneously with scale factor $a(t)=t$, the blob's spikes are the higher dimensional analogue of cusps or garters where a string theory description is locally valid, and the blob's bulk is a negatively curved Einstein manifold. 
%\bf (HMMMMMMMMMMMMMMMMMMMMMMMMMMMMMMMMMMMMM..................................) \rm
%\footnote{We thank T. Levi for suggesting this to us.}

\section{Conclusions}

There are a number of directions it would be interesting to pursue.  Among them would be to study string scattering between different regions of the expanding triniverse or its $\IZ_{2}$ orientifold.  An unoriented string mode propelled towards the oriented part of the space must reflect off the domain wall or transmit into oriented modes, subject to some topological constraints that are relatively simple to work out.  Investigating higher dimensional examples further would be very interesting; for instance, solutions that connect Calabi-Yau manifolds of differing topologies.

The solutions presented in this paper constitute arguably the simplest possible form of time-dependence---a homogeneous expansion of space at constant velocity---and associated with it a very mild type of SUSY breaking (SUSY is broken only globally by boundary conditions, and the scale of the breaking goes to zero at late times as the physical size of the relevant cycles expands).  As such, our metrics are only a baby step away from the kind of time-independent, SUSY configurations typically considered in string theory.   Nevertheless, in \cite{klebanexpand} it was shown that the global constraints on the number of F-theory 7-branes are relaxed by the expansion; here, we have seen that these constructions are sufficiently general to allow a huge variety of effective string compactifications to co-exist simultaneously in one causally connected spacetime.  If such a broad array of possibilities are present even here, we expect more general cosmological solutions to be nearly unconstrained.  Considering that the universe we inhabit is expanding, one must take such possibilities into account.

While the landscape of string theory is not yet well enough understood for us to be able to comment in detail on the broader implications of our findings, it seems likely that the existence of this type of patchwork universe is very relevant to the dynamics of
landscape cosmology.  Much study has been devoted to the implications of the existence of bubbles of different vacua using effective field theory (see e.g. \cite{Czech:2010rg} or \cite{Freivogel:2009it}), but the solutions discussed here present the possibility of collisions of a much more general type---between regions of an 11D manifold effectively described by different 10D string theories or compactifications of such, for example.

\paragraph{Acknowledgements}

The authors would like to thank several people for valuable discussions, including
Nima Arkani-Hamed, Clay Cordova, Frederik Denef, Shmuel Elitzur, Ben Freivogel, Cara Henson,  Albion Lawrence, Tommy Levi, Massimo Porrati, Eliezer Rabinovici, Stephen Shenker, and Leonard Susskind, as well as collaboration
with Ian Swanson during early stages of this project.
SH would like to thank the Center for Cosmology and Particle Physics at
New York University for hospitality while this work was being completed.
The work of SH was supported by the World Premier International Research Center Initiative,
MEXT, Japan, and by a Grant-in-Aid
for Scientific Research (22740153) from the Japan Society for Promotion of
Science (JSPS).   The work of MK is supported by NSF CAREER grant  PHY-0645435.

%%%%%%%%%%%%%%%%%%%%%%%%%%%%%%%%%%%%%%%%%%%%%%%%%%%%%%%%%
%%%%%%%%%%%%%%%%%%%%%%%%%%%%%%%%%%%%%%%%%%%%%%%%%%%%%%%%%
%\renewcommand{\thefigure}{A-\arabic{equation}}
\setcounter{equation}{0}
%%\numberwithin{equation}{section}
%\appendix
%%%%%%%%%%%%%%%%%%%%%%%%%%%%%%%%%%%%%%%%%%%%%%%%%%%%%%%%%
%%%%%%%%%%%%%%%%%%%%%%%%%%%%%%%%%%%%%%%%%%%%%%%%%%%%%%%%%

%%%%%%%%%%%%%%%%%%%%%%%%%%%%%%%%%%%%%%%%%%%%%%%%%%%%%%%%%%%%%%%%%%%%%%%%%%%%%
%
%  APPENDIX -- BEGINS HERE !!
%
%%%%%%%%%%%%%%%%%%%%%%%%%%%%%%%%%%%%%%%%%%%%%%%%%%%%%%%%%%%%%%%%%%%%%%%%%%%%%%%

%%%%%%%%%%%%%%%%%%%%%%%%%%%%%%%%%%%%%%%%%%%%%%%%%%%%%%%%%%%%%%%%%%%%%%%%%%%

\appendix{Hyperbolic nonagons and the trinion}\label{appb}

In this section we will describe the properties of the hyperbolic
trinion.   We choose $\G$ such that
$X$ has the topology of the trinion
(or pair of pants).  This dictates the abstract group structure of $\G$ to
be that of the fundamental group of the trinion, which is the
free group on two generators.

The topology of $X$ determines the abstract structure
of the discrete group $\G$, but the structure of $X$ as a complex manifold
depends on the embedding of $\G$ inside $SL(2,\IR)$.  In particular, we would
like to choose $\G$ such that the three legs of the quotient are
spanned by geodesic "garters" that have a nonzero minimum length.  More
generally, we would like the quotient to be smooth, without cusps or
conical defecits of any kind.  This forces $\G$ to be chosen such that
every element is hyperbolic.  We will refer to this quotient as the \it hyperbolic
trinion. \rm

The hyperbolic trinion discussed here is topologically,
but \it not \rm complex-analytically equivalent to the sphere with three
points removed, which has no moduli at all, whereas the hyperbolic
trinion does have three real moduli, corresponding to the lengths of
the geodesic garters spanning each of the three legs.  For simplicity, we will always consider
the symmetric case, where each of the three garters has
the same length and there is a $\IZ\ll 3$ symmetry rotating
the trinion by $120$ degrees and permuting the three legs cyclically.

We will specify the geometry of the trinion using the description provided by
the uniformization theorem, which allows us to describe any
surface of constant negative curvature as a quotient of $\IH\ll 2$
by a discrete subgroup $\G$.

\subsection{Geometry of hyperbolic space}

We will begin by reviewing some facts about the geometry of
hyperbolic space.  One realization of two-dimensional
hyperbolic space is the
Poincar\'e upper half plane $\IH\ll 2$.  The metric on $\IH\ll 2$ is
\bbb
ds\sqd = {{dx\sqd + dy\sqd }\over{y\sqd}} = {{dz d\zb}\over
{({\rm Im}\cc z)\sqd }} \ .
\eee
This metric has Gaussian curvature $-1$ and Ricci scalar curvature $-2$.
The geodesics in this geometry are circles centered on the
real axis:
\bbb
(x - x\ll c)\sqd + y\sqd = R\sqd\ ,
\eee
for arbitrary $x\ll c$ and $R$.  For a given two points
$x\ll 1$ and $x\ll 2$ the connecting geodesic is given by
\bbb
x\ll c = {{x\ll 1 \sqd - x\ll 2 \sqd + y\ll 1 \sqd -
y\ll 2\sqd}\over{2\cc (x\ll 1 - x\ll 2)}}
\llsk
  R\sqd = \hh \lsqq (x\ll c - x\ll 1)\sqd + (x\ll c - x\ll 2)\sqd
+ y\ll 1 \sqd + y\ll 2 \sqd \rsqq
\eee
and the arc length between them is
\bbb
{\tt dist}(x\ll 1 , y\ll 1 \cc | \cc x\ll 2, y\ll 2) =
{\tt dist}(z\ll 1 \cc | \cc z\ll 2)
\xxn
= {\rm arccosh} \lrdd 1 + {{(x\ll 1 - x\ll 2)\sqd + (y\ll 1 - y\ll 2)\sqd}
\over{2\cc y\ll 1 y\ll 2}} \rrdd
= {\rm arccosh} \lrdd 1 + {{|z\ll 1 - z\ll 2|\sqd}
\over{2\cc {\rm Im}(z\ll 1) {\rm Im}(z\ll 2)}} \rrdd
\ ,
\een{globaldistuhp}
where we use the complex coordinate $z \equiv x + i y$.

Another useful representation of hyperbolic space is
the Poincar\'e disc, $w\in \IC, |w|<1$.  It is
related to the coordinate $z$ by a holomorphic coordinate
transformation
\bbb
w = {{i - z}\over{i + z}} \llsk z = i {{1 - w}\over{1 + w}}.
\eee
In terms of $w$ the metric is
\bbb
{{4 \cc dw d\bar{w}}\over{(1 - |w|\sqd)\sqd}}\ .
\eee
The global distance function in Poincar\'e disc coordinates is
\bbb
{\tt dist}(w\ll 1 \cc | \cc w\ll 2) = {\rm arccosh} \lrdd
1 + {{2 \cc |w\ll 1 - w\ll 2|\sqd}\over{(1 - |w\ll 1 |\sqd)
(1 - |w\ll 2|\sqd)}} \rrdd\ .
\een{globaldistdisc}

A Friedmann-Robertson-Walker cosmology with vanishing stress tensor and
spatial slices given by $\IH\ll 2$ is actually flat Minkowski space.
Start with the FRW metric
\bbb
ds\sqd  = - dt\sqd + t\sqd ds\sqd\ll{\IH\ll 2}
 = - dt\sqd +
{{4 \cc t\sqd \cc dw d\bar{w}}\over{(1 - |w|\sqd)\sqd}}
\eee
and define coordinates
\bbb
X\uu 0 \equiv t \cc {{1 + |w|\sqd}\over{1 - |w|\sqd}}
\xxx
X\uu 1 \equiv {{2 t}\over{1 - |w|\sqd}} \cc {\rm Re}(w)
\llsk
X\uu 2 \equiv {{2t}\over{1 - |w|\sqd}} \cc {\rm Im}(w)
\eee
in which the metric is
\bbb
ds\sqd = - (dX\uu 0)\sqd + (dX\uu 1)\sqd + (dX\uu 2)\sqd \ ,
\eee
now recognizable as the flat metric on Minkowski space.  The inverse
change of coordinates is
\bbb
t = \sqrt{(X\uu 0)\sqd - (X\uu 1)\sqd - (X\uu 2)\sqd} \llsk
w = {{X\uu 0 - \sqrt{(X\uu 0)\sqd - (X\uu 1)\sqd - (X\uu 2)\sqd}}
\over{(X\uu 1)\sqd + (X\uu 2)\sqd}} \cc (X\uu 1 + i X\uu 2)\ .
\eee

\subsection{Isometries of hyperbolic space}

Hyperbolic space has an isometry group $PSL(2,\IR) \simeq SO(2,1)$.
In terms of coordinates on the Poincar\'e upper half plane, the
action of an element $g\in PSL(2,\IR)$ is
\bbb
g: z \mapsto {{p z + q}\over{rz + s}}\ , \llsk g = \lrdd \bm p & q \cr
r & s \em \rrdd\ .
\eee
In terms of the Poincar\'e disc, the action of an element
$\tilde{g}$ is given by
\bbb
\tilde{g} : w\mapsto {{\tilde{p} w + \tilde{q} }\over{\tilde{q}^* w +
\tilde{p}^*}}, \llsk \tilde{g} \equiv \lrdd \bm \tilde{p} & \tilde{q}
\cr \tilde{q}^* & \tilde{p}^* \em \rrdd\ ,
\llsk |\tilde{p}|\sqd - |\tilde{q}|\sqd = 1\ .
\eee
The translation between the two transformations is
\bbb
\lrdd \bm \tilde{p} & \tilde{q} \cr \tilde{q}^* & \tilde{p}^* \em \rrdd
= \lrdd \bm \hh(-p -iq + ir - s)  & \hh(p -iq -ir -s ) \cr
\hh( p +iq + ir - s ) & \hh(-p +iq - ir - s)   \em \rrdd
\xxx
= \lrdd \bm -1 & i \cr 1 & i\em \rrdd \lrdd \bm p & q \cr r & s \em \rrdd
\lrdd \bm -1 & i \cr 1 & i \em \rrdd\uu{-1}\ .
\xxx
\lrdd\bm p & q \cr r & s \em \rrdd = \lrdd \bm
{\rm Re}\cc (\tilde{p} - \tilde{q}) & {\rm Im} \cc (\tilde{p} + \tilde{q})
\cr
- {\rm Im}\cc(\tilde{p} - \tilde{q}) & {\rm Re} \cc (\tilde{p} + \tilde{q})
\em
\rrdd
\xxx
= \lrdd \bm -1 & i \cr 1 & i \em \rrdd\uu{-1}
\lrdd \bm \tilde{p} & \tilde{q} \cr \tilde{q}^* & \tilde{p}^* \em \rrdd
\lrdd \bm -1 & i \cr 1 & i \em \rrdd
\eee
These transformations of course leave $t$ invariant.

In $X\uu\m$ coordinates, the transformations above act as
\bbb
\lsqq \bm X\uu 0 \cr X\uu 1 \cr X\uu 2 \em \rsqq
\mapsto
{\bf L} \cdot
\lsqq \bm X\uu 0 \cr X\uu 1 \cr X\uu 2 \em \rsqq\ ,
\eee
where
$$
{\bf L} \equiv \lrdd \bm
{\bf L}\uu 0{}\ll 0 & {\bf L}\uu 0{}\ll 1 & {\bf L}\uu 0{}\ll 2
\cr
{\bf L}\uu 1{}\ll 0 & {\bf L}\uu 1{}\ll 1 & {\bf L}\uu 1{}\ll 2
\cr
{\bf L}\uu 2{}\ll 0 & {\bf L}\uu 2{}\ll 1 & {\bf L}\uu 2{}\ll 2
\em \rrdd
$$
is given by
${\bf L} \equiv $ \noindent\(\left(
\begin{array}{ccc}
 \text{$\pmacro$} \text{$\pstmacro$}+\text{$\qmacro$} \text{$\qstmacro$} & \text{$\pstmacro$} \text{$\qmacro$}+\text{$\pmacro$}
\text{$\qstmacro$} & i (-\text{$\pstmacro$} \text{$\qmacro$}+\text{$\pmacro$} \text{$\qstmacro$})
\\
 \text{$\pmacro$} \text{$\qmacro$}+\text{$\pstmacro$} \text{$\qstmacro$} & \frac{1}{2} \left(\text{$\pmacro$}^2+\text{$\pstmacro$}^2+\text{$\qmacro$}^2+\text{$\qstmacro$}^2\right)
& \frac{1}{2} i \left(\text{$\pmacro$}^2-\text{$\pstmacro$}^2-\text{$\qmacro$}^2+\text{$\qstmacro$}^2\right)
\\
 -i (\text{$\pmacro$} \text{$\qmacro$}-\text{$\pstmacro$} \text{$\qstmacro$}) & -\frac{1}{2} i \left(\text{$\pmacro$}^2-\text{$\pstmacro$}^2+\text{$\qmacro$}^2-\text{$\qstmacro$}^2\right)
& \frac{1}{2} \left(\text{$\pmacro$}^2+\text{$\pstmacro$}^2-\text{$\qmacro$}^2-\text{$\qstmacro$}^2\right)
\end{array}
\right)\)
\newline

\begin{center}
\noindent\(\pmb{}\)

= \noindent\(\left(
\begin{array}{ccc}
 \text{$|\tilde{p}|\sqd$}+\text{$|\tilde{q}|\sqd$} & 2 ~\text{${\rm Re}(\tilde{p}^*\tilde{q})$} & 2~ \text{${\rm Im}(\tilde{p}^* \tilde{q})$} \\
 2~ \text{${\rm Re}(\tilde{p}\tilde{q})$} & \text{${\rm Re}(\tilde{p}\sqd)$}+\text{${\rm Re}(\tilde{q}\sqd)$} & -\text{${\rm Im}(\tilde{p}\sqd)$}+\text{${\rm Im}(\tilde{q}\sqd)$} \\
 2~ \text{${\rm Im}(\tilde{p}\tilde{q})$} & \text{${\rm Im}(\tilde{p}\sqd)$}+\text{${\rm Im}(\tilde{q}\sqd)$} & \text{${\rm Re}(\tilde{p}\sqd)$}-\text{${\rm Re}(\tilde{q}\sqd)$}
\end{array}
\right)\)
\end{center}
%\newline
\begin{center}
= \noindent\(\left(
\begin{array}{ccc}
 \frac{1}{2} \left(p^2+q^2+r^2+s^2\right) & \frac{1}{2} \left(-p^2+q^2-r^2+s^2\right) & p q+r s \\
 \frac{1}{2} \left(-p^2-q^2+r^2+s^2\right) & \frac{1}{2} \left(p^2-q^2-r^2+s^2\right) & -p q+r s \\
 p r+q s & -p r+q s & q r+p s
\end{array}
\right)\)
\end{center}

\heading{Causal geodesics in FRW coordinates}\label{causgeo}

Now we wish to gain a feel for the causal and dynamical properties of Minkowski
space when viewed in FRW coordinates $(t,w)$ or $(t,z)$.
For instance, we would like to understand when two points, specified
by their FRW coordinates, are causally connected.

Given $(t\lp 1 , z\lp 1)$ and $(t\lp 2, z\lp 2)$, we wish to
write the invariant separation
${\bf S}(X\lp 1 \cc | X\lp 2) \equiv - (X\uu 0 \lp 1 - X\uu 0 \lp 2)\sqd
+ (X\uu 1 \lp 1 - X\uu 1 \lp 2)\sqd + (X\uu 2 \lp 1 - X\uu 2 \lp 2)\sqd$
in terms of their FRW coordinates.  Using the transformation
rules above, we find
\bbb
{\bf S} (t\ll 1 , w\ll 1 \cc | t\ll 2 , w\ll 2)
=
- t\ll 1 \sqd
- t\ll 2 \sqd + 2 t\ll 1 t\ll 2 \cc
\lrdd
1 + {{2 \cc |w\ll 1 - w\ll 2|\sqd}\over{(1 - |w\ll 1 |\sqd)
(1 - |w\ll 2|\sqd)}} \rrdd
\xxx
=
- t\ll 1 \sqd
- t\ll 2 \sqd + 2 t\ll 1 t\ll 2 ~ {\rm cosh}\lrdd {\tt dist}
( w\ll 1 \cc | \cc w\ll 2) \rrdd\ ,
\eee
where ${\tt dist}(w\ll 1 \cc | \cc w\ll 2)$ is the global
distance function on the Poincar\'e disc of unit curvature
radius, as given in equation \rr{globaldistdisc}.
Fixing $t\ll 1 , w\ll 1$ and $w\ll 2$, and taking $0 < t\ll 1 \leq t\ll 2$,
we find that the smallest value of $t\ll 2$ such that the two
points are in causal contact, is
\bbb
t\ll 2 = t\ll 1 \cc \exp{{\tt dist}(w\ll 1 \cc | \cc w\ll 2)}\ .
\eee
  In terms of
the FRW time $t$ and the
coordinate $z$ on the Poincar\'e upper half plane, the
invariant separation between two points is
\bbb
{\bf S}(t\ll 1, z\ll 1 \cc | \cc t\ll 2 , z\ll 2) =
- t\ll 1 \sqd
- t\ll 2 \sqd + 2 t\ll 1 t\ll 2 ~ {\rm cosh}\lrdd {\tt dist}
( z\ll 1 \cc | \cc z\ll 2) \rrdd\ ,
\eee
where ${\tt dist}(w\ll 1 \cc | \cc w\ll 2)$ is the global
distance function on the unit Poincar\'e upper half plane,
given in equation \rr{globaldistuhp}.  So a lightlike signal
sent from point $z\ll 1$ at FRW time $t\ll 1$, arrives at point $z\ll 2$
at FRW time
\bbb
t\ll 2 = t\ll 1 \cc \exp{{\tt dist}(z\ll 1 \cc | \cc z\ll 2)}\ .
\eee
In particular, comoving points (\it i.e.,\rm  points at fixed comoving
coordinates $w\ll i$ or $z\ll i$) never fall out of causal contact with
one another: no matter how distant the points are in $\IH\ll 2$ nor
how late one sends the signal from $z\ll 1$, the signal will always eventually
arrive at $z\ll 2$ after an elapsed time that is exponentially long in the
comoving spacelike distance ${\tt dist}(z\ll 1\cc | \cc z\ll 2)$
on the hyperbolic spatial slice $\IH\ll 2$.  This property
is preserved when we take our quotient
of $\IH\ll 2$ by a discrete group $\G$ to obtain our modified constant-time
spatial slice $X$, which will guarantee that the different
string theories supported in the different regions of $X$
will remain in causal contact with one another for all time.

\subsection{$X$ as a quotient $\IH\ll 2 / \G$}

The hyperbolic trinion is a quotient of $\IH\ll 2$ by a discrete hyperbolic
subgroup $\G$ of its isometry group $SL(2,\IR)$.  For concreteness,
we will now pick a set of generators for $\G$ in the case where
there is a $\IZ\ll 3$ symmetry rotating the trinion's three legs.

Define

 $A \equiv $\noindent\(\pmb{\left(
\begin{array}{cc}
 \frac{1+\gamma^2}{\gamma-\gamma^2} & 1+\gamma  \\
 -\frac{1 \left(1+\gamma^3\right)}{(-1+\gamma)^2 \gamma} & \frac{1+\gamma^2}{-1+\gamma}
\end{array}
\right)}\)

$B \equiv $
\noindent\(\left(
\begin{array}{cc}
 \gamma & 0 \\
 0 & \frac{1}{\gamma}
\end{array}
\right)\)

$C\equiv $
\noindent\(\left(
\begin{array}{cc}
 \frac{1+\gamma^2}{\gamma-\gamma^2} &  1+\frac{1}{\gamma}  \\
 -\frac{ 1+\gamma^3  }{(-1+\gamma)^2} & \frac{1+\gamma^2}{-1+\gamma}
\end{array}
\right)\)

$R\equiv$ \noindent\(\pmb{\left(
\begin{array}{cc}
 \frac{1}{-1+\gamma} & -1 \\
 \frac{1-\gamma+\gamma^2}{(-1+\gamma)^2} & \frac{\gamma}{1-\gamma}
\end{array}
\right)}\)

These matrices have the following properties:
\bi
\item{The third matrix $C$ is the inverse of the product of the
first two, as an element of $PSL(2,\IR)$:
\bbb
A \cdot B \cdot C = - 1\ .
\eee
}
\item{There is a rotation matrix $R$ that permutes the three
matrices $A,B,C$ cyclically acting by conjugation:
\bbb
R\cdot A \cdot R\uu{-1} = B
\xxx
R\cdot B \cdot R\uu{-1} = C\ ,
\xxx
R\cdot C \cdot R\uu{-1} = A\ .
\eee
}
\item{For $\gamma > 0$ and $\gamma \neq 1$ any two of the three generate a discrete hyperbolic
subgroup $\G$ of $SL(2,\IR)$, and the quotient space is a trinion (see Appendix \ref{nona}).
The group $\G$ is also hyperbolic in the ranges
$- \hh (3 - \sqrt {5}) < \gamma < 0$ and $\gamma < -\hh (3 + \sqrt{5})$.
For hyperbolic values
of $\gamma$, the quotient $\IH\ll 2 / \G$ is a smooth surface $X$ of constant
negative curvature.}
\item{Conjugating $A, B$, and $C$ by the $SL(2,\IR)$ matrix
\bbb
J \equiv
\left(
\begin{array}{cc}
 \text{0} & - \frac{1+\gamma}{3 \sqrt{1+\gamma+\gamma^2}}\\
  \frac{3 \sqrt{1+\gamma+\gamma^2}}{1+\gamma} & 0
\end{array}
\right)
\eee
takes $\gamma \rightarrow 1/\gamma$ (that is, $J \cdot A(\gamma) \cdot J^{-1}=A(1/\gamma)$, etc.).  Therefore we restrict our attention to the range $0|\gamma|<1$.}
\item{The homotopy classes of $X$ correspond to
conjugacy classes of $\G$ by inner automorphisms.  The elements
$A, B$ and $C$ all represent distinct conjugacy classes.
The
minimal length curves in each of these homotopy classes are the
geodesics wrapping the narrow point in each of the legs of the
trinion $X$.}
\item{The matrix $R$ acts as an outer automorphism of
$\G$, so it takes $\G$-orbits of points to other $\G$-orbits, and
thus acts as a discrete isometry on
the quotient $X$ of $\IH\ll 2$ by $\G$.
The isometry $R$ permutes
the conjugacy classes corresponding to $A,B$ and $C$, and so
can be thought of geometrically as a $120$ degree rotation that
permutes the three legs of the trinion cyclically.}
\item{The rotation automorphism $R$ has an action on $\IH\ll 2$ with
a single fixed point in the upper half plane $\IH$:
\bbb
R\circ z\ll 0 \upp 1 \equiv z\ll 0 \upp 1
\xxx
z\ll 0\upp 1 \equiv {{(\gamma\sqd - 1) + i (\gamma)~\sqrt{3} \cc (\gamma - 1)\sqd}\over
{2(\gamma\sqd - \gamma + 1) }}\ ,
\eee
where the action of a $PSL(2,\IR)$ transformation
is defined in the usual way:
\bbb
\lrdd \begin{matrix}  p & q \cr r & s \end{matrix} \rrdd \circ z \equiv
{{p\cc z + q}\over{r\cc z + s}}\ .
\eee
Every point in orbit $\G\circ z\ll 0\upp 1$ of $z\ll 0\upp 1$ under
$\G$ is also a fixed point of $R$, up to an action of $\G$, and they
all correspond to a single fixed point of the action of $R$
on $X$.
}
\item{There is a second fixed point of the action of
$R$ on $X$.  For $\gamma > 0$ define
\bbb
z\ll 0\upp 2 \equiv \gamma \cc z\ll 0\upp 1
\eee
and for $\gamma < 0$ define
\bbb
z\ll 0 \upp 2 \equiv \gamma \cc (z\ll 0 \upp 1)^*\ .
\eee
In either case, $z\ll 0\upp 2$ lies in $\IH\ll 2$ and satisfies
\bbb
R\circ z\ll 0 \upp 2 = B\uu{-1}\circ z\ll 0 \upp 2\ .
\eee
Thus $z\ll 0 \upp 2$ is a second fixed point of $R$ up
to $\G$-action, and corresponds to a fixed point of the
action of $R$ on $X$.}
\item{For any hyperbolic value of $\gamma$, the points
$z\ll 0\upp{1,2}$ are the only two fixed points of the action of
$R$ on $X$.  Geometrically, we can understand this by drawing the
axis of $\IZ\ll 3$ symmetry through the center of the trinion, and
marking the two points of intersection, which correspond to
$z\ll 0\upp{1,2}$.}
\item{In addition to the 120 degree rotation $R$ permuting the three
legs of the trinion, there is an independent geometric symmetry
consisting of a 180 degree rotation of one of the three legs, say leg $B$;
call this symmetry $H\ll B$.  For $\gamma > 0$, we can take
$H\ll B$ to be the positive square root of $B$; for $\gamma < 0$ we
can take it to be the positive square root of $-B$.  For any
value of $\gamma$,
\bbb
H\ll B \equiv \lrdd \bm \sqrt{|\gamma|} & 0 \cr 0 & {1\over{\sqrt{|\gamma|}}}
\em \rrdd
\eee
In the range $\gamma > 0$ this matrix satisfies
\bbb
H\ll B \sqd = B
\xxx
H\ll B \cdot B \cdot H\ll B\uu{-1} = B
\xxx
H\ll B \cdot A \cdot H\ll B\uu{-1} = A\uu{-1} \cdot C \cdot A
\xxx
H\ll B \cdot C\cdot  H\ll B\uu{-1} = A\ .
\eee
For $\gamma < 0$ the matrix $H\ll B$ satisfies the same relations, except
that
$H\ll B \sqd = - B$ instead of $+B$; as relations in $PSL(2,\IR)$
therefore, the relations are the same.
Thus $H\ll B$ is an outer automorphism of $\G$ and therefore
acts on $X$.  It acts on conjugacy classes by exchanging
the conjugacy classes of $A$ and $C$, and sending the conjugacy class
of $B$ to itself.  The action of $H\ll B$ is of order $2$ on $X$,
rotating the leg $B$ by 180 degrees, and
permuting the legs $A$ and $C$, as claimed.  Acting with $R$
on $H\ll B$ gives $H\ll A$ and $H\ll C$, whose properties are
obtained from those of $H\ll B$ by permuting $A,B$ and $C$ cyclically.
}
\item{The symmetries $R$ and $H\ll{A,B,C}$ are all holomorphic
symmetries that preserve the orientation of $X$ and have a zero-dimensional
fixed locus.  In the range $\gamma > 0$
there is another set of discrete symmetries that
are antiholomorphic, orientation-reversing and have one-dimensional
fixed loci on $X$.}
\item{To describe the
first of the antiholomorphic symmetries for $\gamma > 0$,
consider a plane slicing all three legs of the
trinion in half, this symmetry should be thought of as a reflection
in that plane.  The action of this symmetry on $\IH\ll 2$
is a complex conjugation combined with the action of a real matrix
$F$.  To find $F$, conider that it must exchange the two
fixed points $z\ll 0\upp {1,2}$, as well as reversing the orientation
of each of the three geodesics around legs $A,B,C$.  Thus we need a
matrix $F$ such that
\bbb
F\cdot A \cdot F\uu{-1} \simeq  A\uu{-1}\xxx
F\cdot B \cdot F\uu{-1} \simeq  B\uu{-1}\xxx
F\cdot C \cdot F\uu{-1} \simeq  C\uu{-1}\ ,
\eee
where the $\simeq$ denotes equality up to inner automorphism,
\it i.e., \rm up to conjugation by an element
of $\G$.  We take
\bbb
F\equiv \lrdd \bm 0 & (\gamma - 1) \sqrt{{\gamma}\over{\gamma\sqd - \gamma + 1}}
\cr {1\over{\gamma - 1}} \cc \sqrt{{\gamma \sqd - \gamma + 1}\over{\gamma}}
& 0 \em \rrdd
\eee
and we have the relations
\bbb
F\sqd = 1
\xxx
F\cdot A \cdot F\uu{-1} = A\uu{-1}
\xxx
F\cdot B \cdot F\uu{-1} = B\uu{-1}
\xxx
F\cdot C \cdot F\uu{-1} = B\cdot C\uu{-1} \cdot B\uu{-1}\ ,
\eee
so the action of $F$ takes each of the conjugacy classes of the three
generators to its inverse.  This is the correct action for
a reflection through the plane slicing all three legs in half.
The linear operation $F$ is not an isometry of $\IH\ll 2$,
but it is when combined with the complex conjugation operation $*$, and
the comibned operation ${\bf F} \equiv F \cdot * = * \cdot F$ has the same
commutation relations as does $F$, since $*$ commutes with
$A,B$ and $C$, as well as with $H\ll{A,B,C}$ and $R$.  So
\bbb
{\bf F}: z \mapsto {{\gamma \sqd - \gamma + 1 }\over{\gamma\cc
(\gamma - 1)\sqd \cc \bar{z}}}
\eee
acts as a discrete antiholomorphic isometry of order 2 on $X$.
}
\item{The situation with respect to antiholomorphic symmetries is
slightly different in the range $\gamma < 0$.  For $\gamma < 0$ we
can define an $F$-matrix
\bbb
F\equiv \lrdd \bm 0 & (\gamma - 1) \sqrt{{(-\gamma)}\over{\gamma\sqd - \gamma + 1}}
\cr - {1\over{\gamma - 1}} \cc \sqrt{{\gamma \sqd - \gamma + 1}\over{(-\gamma)}}
& 0 \em \rrdd
\eee
which has the same relations with $A,B$ and $C$ as does
the $F$-matrix for the case $\gamma > 0$, except that $F\sqd = -1$
instead of $+1$.  The action of $F$ takes $z$ to ${1\over{\gamma \cc z}}$
as it does in the case $\gamma > 0$.  But in the case $\gamma < 0$ this
action takes the upper half plane to itself, without an
additional complex conjugation on $z$.  Thus the matrix $F$ corresponds
to an additional holomorphic, rather than an antiholomorphic
symmetry.  This symmetry has at least one fixed point
in $\IH\ll 2$, given by $z = {{+i}\over{\sqrt{-\gamma}}}$.
The familiar $\IZ\ll 3$-symmetric trinion has no
such holomorphic symmetry that reverses the orientations of
the geodesics around each leg.}
\item{Based on the absence of such a symmetry in the geometry of
interest, we shall henceforth ignore the case $\gamma < 0$, and
we will assume $\gamma > 0$ going forward.\footnote{Using the technique in Appendix \ref{nona}, one can
verify that $\gamma<0$ corresponds to a torus with either a single circular hyperbolic boundary component cut out ($0>\gamma>-(3 - \sqrt{5})/2$) or
to a torus with a single conical defect ($\gamma<-(3 - \sqrt{5})/2$).}}
\item{For $\gamma > 0$ there are three additional antiholomorphic
symmetries ${\bf K}\ll{A,B,C} \equiv K\ll{A,B,C} \cdot *
= * \cdot K\ll{A,B,C}$, where $K\ll{A,B,C}$ are three matrices
in $SL(2,\IR)$ that we now describe.  Starting with $K\ll B$, we
take
\bbb
K\ll B \equiv \lrdd \bm  0 & {{\gamma - 1}\over{\sqrt{\gamma\sqd - \gamma + 1}}}
\cr {{\sqrt{\gamma\sqd - \gamma + 1}}\over{\gamma - 1}} & 0 \em \rrdd
\eee
which has the properties
\bbb
K\ll B\sqd = 1
\xxx
K\ll B\cdot B \cdot K\ll B\uu{-1} = B\uu{-1}
\xxx
K\ll B\cdot A \cdot K\ll B\uu{-1} = C\uu{-1}
\xxx
K\ll B\cdot C \cdot K\ll B\uu{-1} = A\uu{-1}
\eee

The combination ${\bf K}\ll B \equiv K\ll B \cdot * = * \cdot K\ll B$
is an isometry of $\IH$ that permutes conjugacy classes of
$\G$, and acts on $z$ as
\bbb
{\bf K}\ll B : \llsk z \mapsto {{\gamma\sqd - \gamma + 1}\over{(\gamma - 1)\sqd
\cc \zb}}
\eee
The matrix $K\ll B$ is important to us, because it is the geometric
symmetry by which we will orientifold in order to obtain the
interpolating solution between oriented and unoriented string theories.
Conjugating $K\ll B$ by $R$, of course, yields two more matrices $K\ll C$ and
$K\ll A$ that have properties corresponding to those of $K\ll B$, with
the matrices $A,B$ and $C$ permuted cyclically.
}
\ei
The quotient has no continuous isometries, though there are
approximate continuous isometries in the middle of the legs.
The $B$-leg has an approximate continuous isometry given
by
\bbb
\co\ll B \equiv
 \lrdd z\pp\ll z + \zb\pp\ll\zb \rrdd
=
{{\pp}\over{\pp{\cc \rm ln}
\cc (r)}}\ .
\eee
Then the identification $B: z\mapsto \gamma\sqd \cc z$ can be realized
as $\exp{2a\cc \co\ll B}$, where $a = {\rm ln}\gamma$.  We'd like to
pick a compact coordinate $\chi$ with fixed identifications such that
$B$ acts as $\chi\to \chi + 1$, so
\bbb
\chi = {{1}\over{2\cc \rm ln \gamma}}\cc {\rm ln}~ r\ ,
\eee
and
\bbb
\pp\ll\chi = 2\cc \ln\gamma  \cc \pp\ll{{\rm ln}~r} =2\cc \ln\gamma  \cc \co\ll B .
\eee

$\co\ll B$ itself does not act on $X$ because it is not
invariant under $\G$.  However $\co\ll B$ does extend, as a vector field, to an
open set containing the garter of the B-leg; the breaking of the
isometry generated by $\co\ll B$ is therefore nonperturbative in
the ratio of $c\cc H\uu{-1}$ to $\bfr\ll 0$.

We change to the coordinate system
\bbb
r = \exp{2\cc {\rm ln}(\gamma) \cc \chi} ,
\llsk \th = 2 \cc {\rm tan}\uu{-1} (e\uu u)\ ,
\xxx
\chi = {{1}\over {2\cc \rm ln}(\gamma)} {\rm ln} \cc r\ ,
\llsk u = {\rm ln}\cc {\rm tan}\cc (\th / 2)\ .
\xxx
{\rm sin}(\th) = {{2 e\uu u}\over{1 + e\uu{2u}}} = {{d\th}\over{du}}
\eee
in which the Poincar\'e metric takes the form
\bbb \label{legb}
du\sqd +
4\cc {\rm ln}\sqd(\gamma) \cc {\rm cosh}\sqd(u) \cc d\chi\sqd\ .
 \eee
We will refer to this system as \it leg-B coordinates, \rm for it is
particularly well adapted to describe phenomena localized in the
leg of the trinion whose ``garter'' geodesic is parametrized by
$u=0, \chi \in [0,1)$.  Each leg has such a coordinate system, in which
the compact vector field translating the circular fibers acts as an
approximate isometry whose breaking is suppressed to all orders in $ {{r\ll 0 H}\over c}$.

\subsection{Hyperbolic nonagons} \label{nona}

To construct the quotient space $\IH_2/\Gamma$ given the generators above, we wish to find a fundamental domain of the subgroup $\Gamma$ in either the upper half-plane or Poincare disk.  We will see that the simplest representation of the fundamental domain is a nonagon with certain identifications on its edges.  Three of the nonagon's edges are sections of the real axis (or the boundary of the disk $|z|=1$), and six are geodesic arcs.  For $x>0$ the endpoints of each of the three real axis sections are identified, and so constitute the three circles at the hyperbolic boundary of the trinion.\footnote{For $x<0$ they are identified in such a way that together they form a single circle.  In that case the topology is different---it is a torus with a disk cut out.  For sufficiently negative $x$ the circle pinches off into a cusp, and then into a conical defect (so that the space is topologically a torus with a conical defect).  We will not make use of the regime $x<0$, although it could be of interest for studying topologically non-trivial ``BTZ black holes'' or for other purposes.}

To demonstrate this we first need a set of rules for constructing the fundamental domain given a generating set for some finitely-generated subgroup $\Gamma$.  We will construct the canonical Fricke polygon (the simplest representation of the fundamental domain) for $\Gamma$ following the procedure found in \cite{keen}.

In the case where all the generators in the generating set of $\Gamma$ are hyperbolic, one begins by identifying the {\it isometric circles}  \cite{ford}  of each generator and its inverse in the Poincare disk representation of $\IH_2$.  For an $SL(2,\IR)$ generator $A$ that maps $z \rightarrow z'=(az+b)/(cz+d)$, $c \neq 0$, the isometric circle is the geodesic  arc defined by $|cz+d|=1$ (the name ``isometric circle'' follows from the fact that the Jacobian of the transformation is $(cz+d)^2$, so this circle is the set of points on which the action of the generator preserves the determinant of the metric and doesn't ``stretch'' the space).
The isometric circle is not fixed under the action of $A$; rather, it is mapped into the isometric circle of $A^{-1}$ \cite{ford}.

The canonical Fricke polygon of a group generated by hyperbolic transformations is simply the region to the interior of all the isometric circles of the generators in the generating subgroup and their inverses (where the ``interior" of an isometric circle is defined as the side containing the isometric circle of the inverse generator).  One imposes on the boundaries of this region the identifications due to the action of each respective generator ({\it i.e.}, each generator's isometric circle is identified with the isometric circle of its inverse).  In our case this is simplest to visualize in the Poincare disk representation of $H_{2}$; see Fig. \ref{nonagon}.

Another important geodesic defined by $A$ (or $A^{-1}$) is its {\it axis}.  The axis is the geodesic arc that connects the two fixed points of $A$ (which for a hyperbolic generator are two distinct points on the hyperbolic boundary).  The axis intersects both the isometric circle of $A$ and the isometric circle of $A^{{-1}}$ at right angles.  Because these two isometric circles are identified under $\Gamma$, the section of the axis that connects them forms a closed regular geodesic, and in fact is the garter corresponding to $A$ (the minimum length geodesic around leg A of the trinion).

From this construction one can easily compute the geodesic length of the garter and the geodesic distance from a point on the garter to the $\IZ_{3}$ symmetric point.  The distance around the garter is trivial to calculate using the upper half-plane representation, in which one of the generators takes the form $a=\gamma, b=0, c=0, d=1/\gamma$, so that $z \rightarrow \gamma^{2} z$. The result is that the distance is 
\bbb
{\tt dist} = 2 |{\rm ln}\cc \g|\ .
\eee

\end{document}